\documentstyle[preprint,eqsecnum,prd,aps,epsf]{revtex}
\begin{document}
\draft 

\tighten   

\newcommand\mup{\mu}      
\newcommand\mutrue{\mu_t} 
\newcommand\mub{b}        
\newcommand\mubest{\mu_{\rm best}}
\newcommand\mubesti{\mu_{{\rm best}_i}}
\newcommand\Tbest{T_{\rm best}}
\newcommand\notitle{}

\draft
\preprint{HUTP-97/A096}
\title{A Unified Approach to the Classical\\
Statistical Analysis of Small Signals}
\author{Gary J. Feldman\cite{bylineGF}}
\address{Department of Physics, Harvard University, Cambridge, MA 02138}
\author{Robert D. Cousins\cite{bylineRC}}
\address{Department of Physics and Astronomy, University of California,
Los Angeles, CA 90095}
\date{\today}
\maketitle
\begin{abstract}

We give a classical confidence belt construction which unifies the
treatment of upper confidence limits for null results and two-sided
confidence intervals for non-null results.  The unified treatment
solves a problem (apparently not previously recognized) that the
choice of upper limit or two-sided intervals leads to intervals which
are not confidence intervals if the choice is based on the data.  We
apply the construction to two related problems which have recently
been a battle-ground between classical and Bayesian statistics:
Poisson processes with background, and Gaussian errors with a bounded
physical region.  In contrast with the usual classical construction
for upper limits, our construction avoids unphysical confidence
intervals.  In contrast with some popular Bayesian intervals, our
intervals eliminate conservatism (frequentist coverage greater than
the stated confidence) in the Gaussian case and reduce it to a level
dictated by discreteness in the Poisson case.  We generalize the
method in order to apply it to analysis of experiments searching for
neutrino oscillations.  We show that this technique both gives correct
coverage and is powerful, while other classical techniques that have
been used by neutrino oscillation search experiments fail one or both
of these criteria.
\end{abstract}


\narrowtext

\section{Introduction}

Classical confidence intervals are the traditional way in which high
energy physicists report errors on results of experiments.
Approximate methods of confidence interval construction, 
in particular the likelihood-ratio method, are
often used in order to reduce computation.  When these approximations
are invalid, true confidence intervals can be obtained using the
original (defining) construction of Neyman \cite{Neyman}.  In recent
years, there has been considerable dissatisfaction with the usual
results of Neyman's construction for upper confidence limits, in
particular when the result is an unphysical (or empty set) interval.
This dissatisfaction led the Particle Data Group (PDG) \cite{PDG} to
describe procedures for Bayesian interval construction in the
troublesome cases: Poisson processes with background, and Gaussian
errors with a bounded physical region.

In this paper, we use the freedom inherent in Neyman's construction in
a novel way to obtain a unified set of classical confidence intervals
for setting upper limits and quoting two-sided confidence intervals.
The new element is a particular choice of ordering, based on
likelihood ratios, which we substitute for more common choices in
Neyman's construction.  We then obtain confidence intervals which are
never unphysical or empty. Thus they remove an original motivation for
the description of Bayesian intervals by the PDG.

Moreover, we show below that commonly quoted confidence intervals are
wrong {\em more} than allowed by the stated confidence {\em if} (as is
typical) one uses the experimental data to decide whether to consult
confidence interval tables for upper limits or for central confidence
intervals.  In contrast, our unified set of confidence intervals
satisfies (by construction) the classical criterion of frequentist
coverage of the unknown true value.  Thus the problem of wrong
confidence intervals is also solved.

Our intervals also effectively decouple the calculation of intervals
from the test of goodness-of-fit, which is desirable but in fact not
the case for traditional classical upper limit calculations.

After developing the new intervals for the two prototypical 1-D
problems, we generalize them for use in the analysis of experiments
searching for neutrino oscillations, continuing to adhere to the
Neyman construction.

In Sec.~\ref{sec-review}, we review and contrast Bayesian and
classical interval construction.  In Sec.~\ref{sec-exam}, we review
the troublesome cases of Poisson processes with background and
Gaussian errors with a bounded physical region.  We introduce the
unifying ordering principle in Sec.~\ref{sec-order}, and apply it to
the previously discussed problems.  In Sec.~\ref{sec-neut}, we
generalize the method for use in neutrino oscillation searches, and
compare it to other classical methods.  Finally, in
Sec.~\ref{sec-excess}, we introduce an additional quantity helpful in
describing experiments which observe less background than expected.
We conclude in Sec.~\ref{sec-conclude}.

We adopt the following notation: the subscript $t$ on a parameter
means the unknown true value; the subscript 0 means a particular
measured value obtained by an experiment.  Thus, for example, $\mup$
is a parameter whose true value $\mutrue$ is unknown; $n_0$ is the
particular result of an experiment which measures the number of events
$n$.  For most of our discussion, we use for illustration 90\%
C.L. confidence intervals on a single parameter $\mup$.  The
confidence level (C.L.)  is more generally called $\alpha$.

\section{Bayesian and Classical Interval Constructions}
\label{sec-review}
\subsection{Bayesian Intervals}
\label{sec-bayes}

Although our approach is classical, it is worthwhile to review
Bayesian intervals since we find that misconceptions about classical
intervals can have their roots in misinterpreting them as Bayesian
intervals.  For advocacy of Bayesian intervals in high energy physics,
see, for example, Refs.\ \cite{Helene,ProsperNIMPRD}.

Suppose that we wish to make an inference about a parameter $\mup$
whose true value $\mutrue$ is unknown.  Assume that we do this by
making a single measurement of an observable $x$ such that the
probability density function (pdf) for obtaining the value $x$ depends
on the unknown parameter $\mup$ in a known way: we call this pdf
$P(x|\mup)$
\cite{comma-vertical}.  (Note that $x$ need not be a measurement
of $\mup$, though that is often the case; $x$ just needs to be 
some observable whose pdf depends on $\mup$.)

Now suppose that the single measurement of $x$ yields the value $x_0$.
One substitutes this value of $x$ into $P(x|\mup)$ to obtain
$P(x_0|\mup)$, known as the likelihood function, which we denote
${\cal L}(x_0|\mup)$.

The Bayesian deems it sensible to speak of pdf's for the unknown
$\mutrue$; these pdf's represent degree of belief about $\mutrue$. One
makes inferences using the ``posterior'' pdf, which is the conditional
pdf $P(\mutrue|x_0)$ for the unknown $\mutrue$, {\em given} the result
$x_0$ of the measurement.  It is related to ${\cal L}$ by applying
Bayes's Theorem.  Bayes's Theorem in classical probability says that
the probability that an element is in both sets $A$ and $B$ is
$P(A|B)\,P(B) = P(B|A)\,P(A)$.  Bayesians apply this to pdf's for
$\mutrue$, obtaining
\begin{equation} 
\label{eqn-bayes}
P(\mutrue|x_0) = {\cal L}(x_0|\mutrue) \ P(\mutrue)/P(x_0).
\end{equation} 
Typically the denominator is just a normalization constant, so the
major issue is what to use for $P(\mutrue)$, which is called the
``prior'' pdf.  For the moment we assume that one has the prior pdf,
so that then one has the posterior pdf.

A Bayesian interval $[\mu_1,\mu_2]$ corresponding to a confidence
level $\alpha$ can be constructed from the posterior pdf by requiring
\begin{equation} 
\label{eqn-bayes_int}
\int_{\mu_1}^{\mu_2} P(\mutrue|x_0) d\mutrue = \alpha.
\end{equation} These
intervals are more properly called `` credible intervals'', although
the phrase ``Bayesian confidence intervals'' is also used \cite{Zacks}.
Note
that there is freedom in the {\em choice} of $\mu_1$ depending on
whether one desires an upper limit, lower limit, central interval,
etc.

We believe that for making decisions, this Bayesian description of
inference is probably how many scientists do (and should) think, and
that the prior pdf one uses is typically the {\em subjective} prior.
One person's subjective prior incorporates all of that person's
personal beliefs as well as the results of previous experiments.
Thus, values of $\mup$ which contradict well-founded theoretical
thinking are (properly) given a low prior \cite{Anderson}.

There have been long-standing attempts to take the subjectivity out of
the prior pdf, in order to have an ``objective'' Bayesian interval.
One attempts to define a prior pdf which represents prior ignorance,
or which is ``non-informative''.  The naive choice of uniform prior is
not well-defined for a continuous variable, since one must specify in
what metric the prior is uniform; this is just as hard as specifying
the functional form in a particular given metric.  For a parameter
$\mu$ which is restricted to $[0,\infty]$, a common non-informative
prior in the statistics literature \cite{Jeffreys,Jaynes} is $P(\mutrue) =
1/\mutrue$, which corresponds to a uniform prior for $\ln\mutrue$.  An
alternative \cite{Kendall-vol2,Jeffreys46} for the Poisson mean is
$P(\mutrue) = 1/\sqrt{\mutrue}$.  In contrast, the PDG recommendation is
equivalent to using a prior which is uniform in $\mutrue$. This
recommendation has no basis that we know of in Bayesian theory.  It is
based on the desire to have intervals which are conservative (see
below) and somewhat robust from a frequentist (anti-Bayesian) point of
view.

In our view, the attempt to find a non-informative prior within
Bayesian inference is misguided.  The real power of Bayesian inference
lies in its ability to incorporate ``informative'' prior information,
not ``ignorance''.  The interpretation of Bayesian intervals based on
uniform priors is vague at best, since they may bear no relation
either to subjective Bayesian intervals of a typical scientist, or to
classical confidence intervals which are probability statements based
only on $P(x|\mup)$.

\subsection{Classical Confidence Intervals}
\label{sec-class}

Neyman's original ``confidence intervals'' \cite{Neyman} completely
avoid the concept of pdf's in $\mutrue$, and hence have no troublesome
prior.  They are limited to statements derived from $P(x|\mup)$; in
our experience this can lead to misinterpretations by those who
mistakenly take them to be statements about $P(\mutrue|x_0)$.  We
believe that, compared to Bayesian intervals with an ``objective
prior'', confidence intervals provide the preferred option for
publishing numerical results of an experiment in an objective way.
However, it is critical not to interpret them as Bayesian intervals,
i.e., as statements about $P(\mutrue|x_0)$.  Rather, a confidence interval
$[\mu_1,\mu_2]$ is a {\em member of a set}, such that the set
has the property that
\begin{equation}
\label{eqn-cover}
P(\mup \in [\mu_1,\mu_2]) = \alpha.
\end{equation}
Here $\mu_1$ and $\mu_2$ are functions of the measured $x$,
and Eq.~\ref{eqn-cover} refers to the {\em
varying} confidence intervals $[\mu_1,\mu_2]$ from an ensemble of
experiments with {\em fixed} $\mup$. For a set of confidence
intervals,  Eq.~\ref{eqn-cover} is true for every allowed $\mup$.
Thus, in particular, the intervals contain the {\em fixed unknown}
$\mutrue$ in a fraction $\alpha$ of experiments.  This is entirely
different from the Bayesian statement that the degree of belief that
$\mutrue$ is in $[\mu_1,\mu_2]$ is $\alpha$.

If Eq.\ (\ref{eqn-cover}) is satisfied, then one says that the
intervals ``cover'' $\mup$ at the stated confidence, or
equivalently, that the set of intervals has the correct ``coverage''.
If there is any value of $\mup$ for which $P(\mup \in
[\mu_1,\mu_2]) < \alpha$, then we say that the intervals
``undercover'' for that $\mup$.  Significant undercoverage for any
$\mup$ is a serious flaw.  If there is any value of $\mup$ for which
$P(\mup \in [\mu_1,\mu_2]) > \alpha$, then we say that the
intervals ``overcover'' for that $\mup$.  A set of intervals is
called ``conservative'' if it overcovers for some values of $\mup$
while undercovering for no values of $\mup$.  Conservatism, while not
generally considered to be as serious a flaw as undercoverage, comes
with a price: loss of power in rejecting false hypotheses.

\begin{figure}
\begin{center}
\leavevmode
\epsfxsize=15cm
\epsfbox{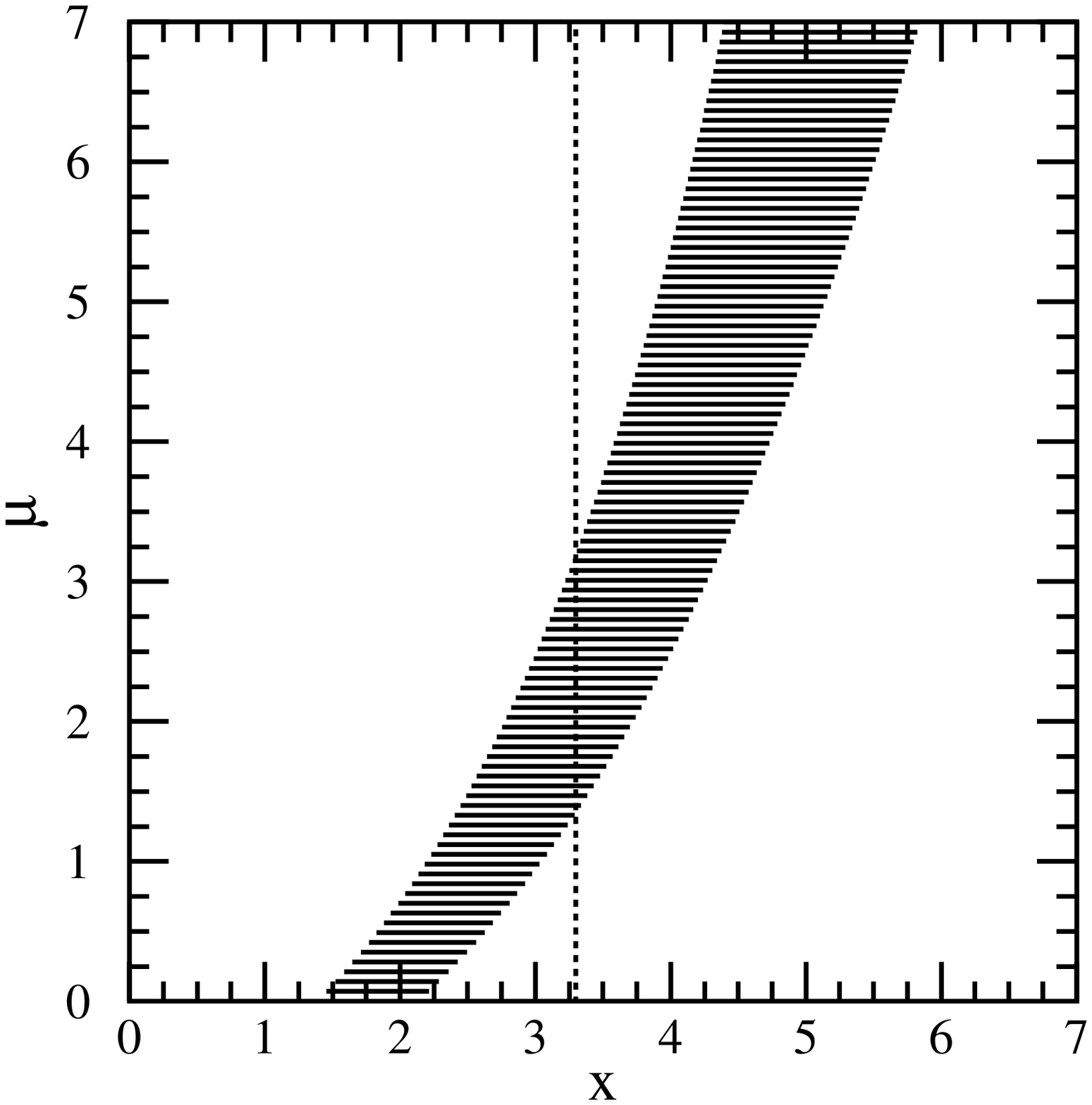}
\end{center}
\caption{A generic confidence belt construction and its use.
For each value of $\mup$, one draws a horizontal acceptance interval
$[x_1,x_2]$ such that $P(x\in [x_1,x_2]\,|\mup) = \alpha$.  Upon
performing an experiment to measure $x$ and obtaining the value $x_0$,
one draws the dashed vertical line through $x_0$.  The confidence
interval $[\mu_1,\mu_2]$ is the union of all values of $\mup$ for
which the corresponding acceptance interval is intercepted by the
vertical line.}
\label{fig-example-belt}
\end{figure}

Our confidence intervals require the full power of Neyman's
construction, which for one measured quantity and one unknown
parameter is called the method of ``confidence belts''
\cite{Kendall-vol2,Eadie}.  Figure \ref{fig-example-belt} illustrates
such a construction on a graph of the parameter $\mup$ vs.\ the
measured quantity $x$.  For each value of $\mup$, one examines
$P(x|\mup)$ along the horizontal line through $\mup$.  One selects an
interval $[x_1,x_2]$ which is a subset of this line such that
\begin{equation}
\label{eqn-acc-region}
P(x\in [x_1,x_2]\,|\mup) = \alpha.
\end{equation}
Such intervals are drawn as horizontal line segments on
Fig.~\ref{fig-example-belt}, at representative values of $\mup$.  We
refer to the interval $[x_1,x_2]$ as the ``acceptance region'' or the
``acceptance interval'' for that $\mup$.  In order to specify uniquely
the acceptance region, one must {\em choose} auxiliary criteria.  One
has total freedom to make this choice, {\em if the choice is not
influenced by the data $x_0$}. The most common choices are:
\begin{equation}
\label{eqn-upper}
P(x<x_1|\mup) = 1-\alpha,
\end{equation}
which leads to ``upper confidence limits'' 
(which satisfy $P(\mup>\mu_2)=1-\alpha$); and
\begin{equation} 
\label{eqn-central}
P(x<x_1|\mup) = P(x>x_2|\mup) = (1-\alpha)/2,
\end{equation}
which leads to ``central confidence intervals''
(which satisfy $P(\mup<\mu_1)=P(\mup>\mu_2)=(1-\alpha)/2$).  
For these choices,
the full confidence belt construction is rarely mentioned, since a
simpler explanation suffices when one specifies $P(x<x_1|\mup)$ and
$P(x>x_2|\mup)$ {\em separately}.  For more complicated choices which
still satisfy the more general specification of Eq.\
(\ref{eqn-acc-region}), an ordering principle is needed to specify
which $x$'s to include in the acceptance region.  We give our ordering
principle in Sec.~\ref{sec-order}.

The construction is complete when horizontal acceptance intervals are
drawn for each value of $\mup$.  Upon performing an experiment to
measure $x$ and obtaining the value $x_0$, one draws a vertical line
(shown dashed in Fig.~\ref{fig-example-belt}) through $x_0$ on the
horizontal axis.  The confidence interval is the union of all values
of $\mup$ for which the corresponding horizontal interval is
intercepted by the vertical line; typically this is a simply connected
interval $[\mu_1,\mu_2]$.  When displayed in texts, typically only the
endpoints of the intervals are drawn, which collectively form the
``confidence belt''.

By construction, Eq.\ (\ref{eqn-cover}) is satisfied for all $\mup$;
Hence it is satisfied for  $\mutrue$, whose value is fixed but unknown.

\section{Examples of Classical Intervals}
\label{sec-exam}
\subsection{Gaussian with Boundary at Origin}

\begin{figure}
\begin{center}
\leavevmode
\epsfxsize=15cm
\epsfbox{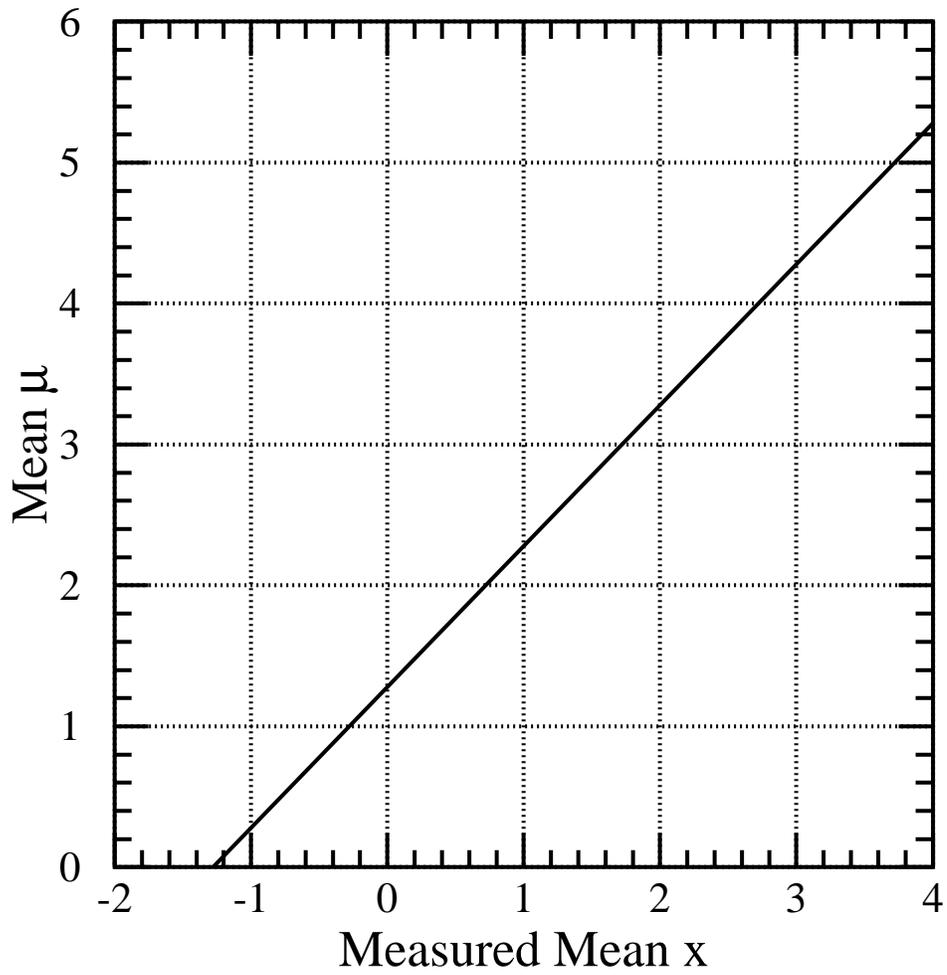}
\end{center}
\caption{Standard confidence belt for 90\% C.L. upper limits
for the mean of a Gaussian, in units of the rms deviation. 
The second line in the belt is at $x=+\infty$.}
\label{fig-std-gauss-ul}
\end{figure}

\begin{figure}
\begin{center}
\leavevmode
\epsfxsize=15cm
\epsfbox{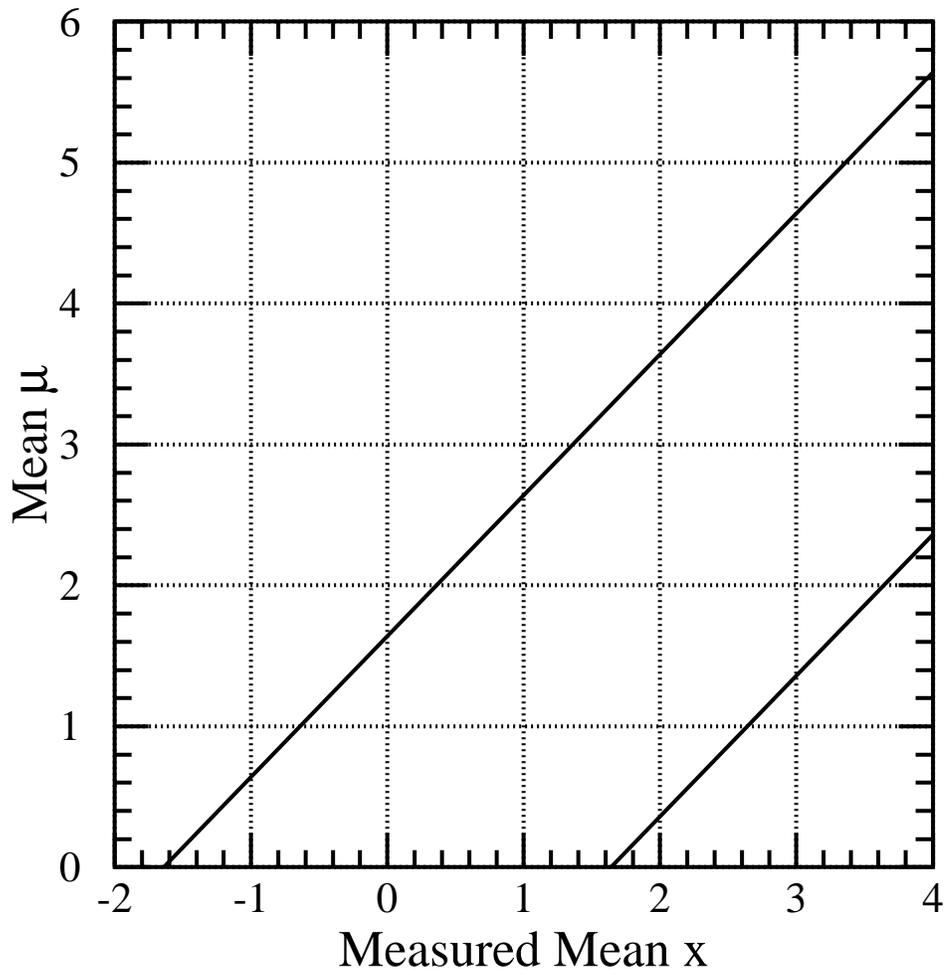}
\end{center}
\caption{Standard confidence belt for 90\% C.L. central confidence
intervals for the mean of a Gaussian, in units of the rms deviation.}
\label{fig-std-gauss-central}
\end{figure}

Figures \ref{fig-std-gauss-ul} and \ref{fig-std-gauss-central} show
standard confidence belts (for upper limits and central intervals,
respectively) when the observable $x$ is simply the measured value of
$\mu$ in an experiment with a Gaussian resolution function with known
fixed rms deviation $\sigma$, set here to unity.  I.e.,
\begin{equation}
\label{eqn-gauss}
P(x|\mup)  = {1\over\sqrt{2\pi}}\exp(-(x-\mup)^2/2).
\end{equation}
We consider the interesting case where only non-negative values for
$\mup$ are physically allowed (for example, if $\mup$ is a
mass).  Thus, the graph does not exist for $\mup<0$.

Although these are standard graphs, we believe that common use of them
is not entirely proper.  Fig.\ \ref{fig-std-gauss-ul}, constructed
using Eq.\ \ref{eqn-upper}, is appropriate for experiments {\em when
it is determined before performing the experiment that an upper limit
will be published}.  Fig.\ \ref{fig-std-gauss-central}, constructed
using Eq.\ \ref{eqn-central}, is appropriate for experiments {\em when
it is determined before performing the experiment that a central
confidence interval will be published}.  However, it may be deemed
more sensible to decide, {\em based on the results of the experiment},
whether to publish an upper limit or a central confidence interval.

Let us suppose, for example, that Physicist X takes the following
attitude in an experiment designed to measure a small quantity: ``If
the result $x$ is less then $3\sigma$, I will state an upper limit
from the standard tables.  If the result is greater than $3\sigma$, I
will state a central confidence interval from the standard tables.''
We call this policy ``flip-flopping'' based on the data.  Furthermore,
Physicist X may say, ``If my measured value of a physically positive
quantity is negative, I will pretend that I measured zero when quoting
a confidence interval'', which introduces some conservatism.

\begin{figure}
\begin{center}
\leavevmode
\epsfxsize=15cm
\epsfbox{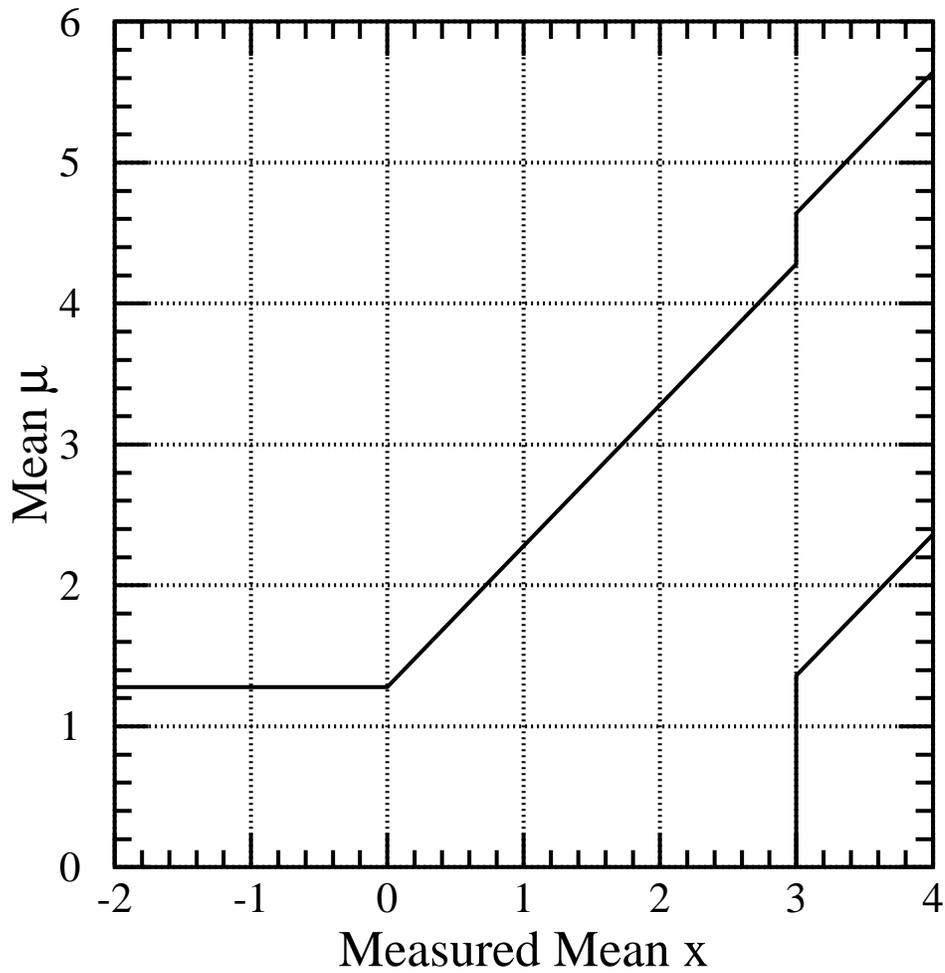}
\end{center}
\caption{Plot of confidence belts implicitly used for
90\% C.L. confidence intervals (vertical intervals between the belts)
quoted by flip-flopping Physicist X, described in the text.  They are
not valid confidence belts, since they can cover the true value at a
frequency less than the stated confidence level.  For
$1.36<\mup<4.28$, the coverage (probability contained in the
horizontal acceptance interval) is 85\%.}
\label{fig-gauss-flipflop}
\end{figure}

We can examine the effect of such a flip-flopping policy by displaying
it in confidence-belt form as shown in Fig.\ \ref{fig-gauss-flipflop}.
For each value of measured $x$, we draw at that $x$ the vertical
segment $[\mu_1,\mu_2]$ that Physicist X will quote as a confidence
interval.  Then we can examine this collection of vertical
confidence intervals to see what horizontal acceptance intervals it
implies.  For example, for $\mup=2.0$, the acceptance interval has
$x_1=2-1.28$ and $x_2=2+1.64$.  This interval only contains 85\% of
the probability $P(x|\mup)$.  Thus Eq.~(\ref{eqn-acc-region}) is
not satisfied. Physicists X's intervals {\em undercover} for a
significant range of $\mup$: they are {\em not} confidence
intervals or conservative confidence intervals.

Both Figs. \ref{fig-std-gauss-ul} and \ref{fig-std-gauss-central} {\em
are} confidence intervals when used appropriately, i.e., without
flip-flopping.  However, the
result is unsatisfying when one measures, for example, $x=-1.8$.  In
that case, one draws the vertical line as directed and finds that
the confidence interval is the empty set!  (An alternative way of
expressing this situation is to allow non-physical $\mup$'s when
constructing the confidence belt, and then to say that the confidence
interval is entirely in the non-physical region.  This requires 
knowing $P(x|\mup)$ for non-physical $\mup$, which can raise
conceptual difficulties.)  When this situation
arises, one {\em knows} that one is in the ``wrong'' 10\% of the
ensemble quoting 90\% C.L. intervals.  One can go ahead and quote the
wrong result, and the ensemble of intervals will have the proper
coverage.  But this is not very comforting.

Both problems of the previous two paragraphs are solved by the
ordering principle which we give in Sec.\ \ref{sec-order}.

\subsection{Poisson with Background}

\begin{figure}
\begin{center}
\leavevmode
\epsfxsize=15cm
\epsfbox{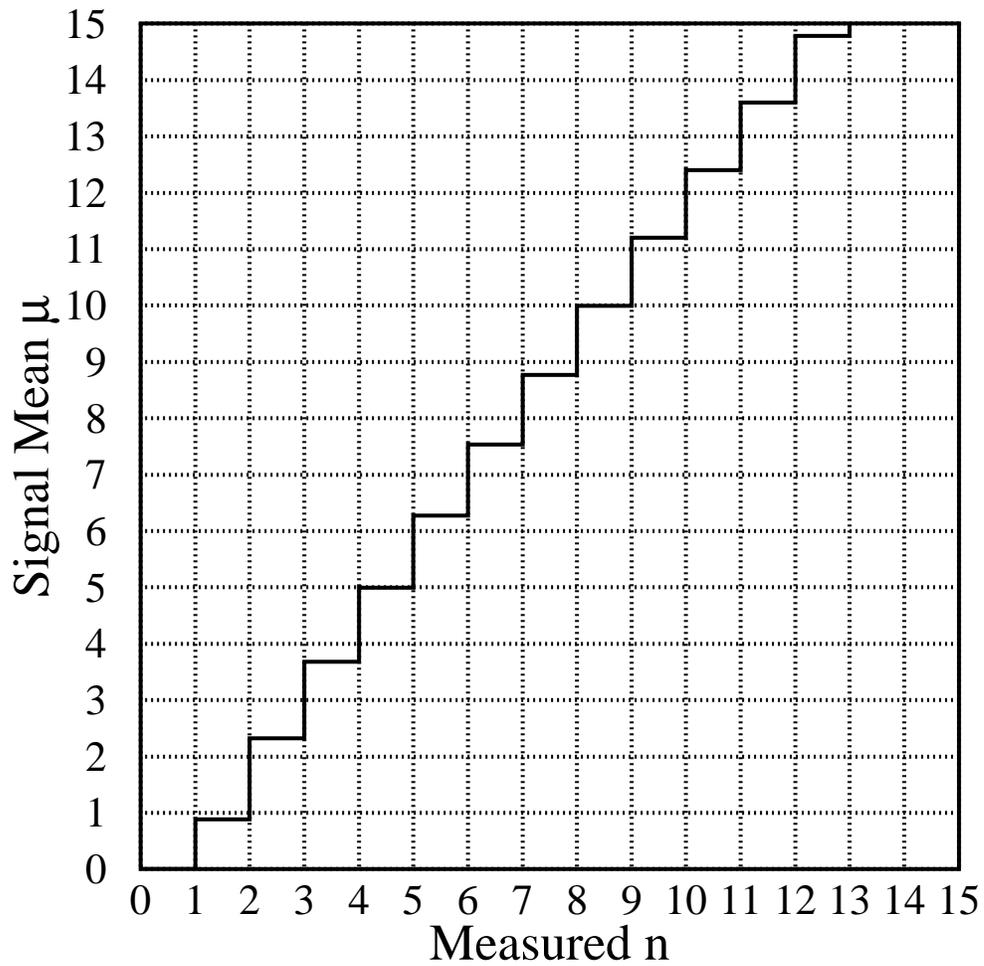}
\end{center}
\caption{Standard confidence belt for 90\% C.L. upper limits,
for unknown Poisson signal mean $\mup$ in the presence of Poisson
background with known mean $\mub=3.0$.  The second line in the belt
is at $n=+\infty$.}
\label{fig-std-pois-ul}
\end{figure}

\begin{figure}
\begin{center}
\leavevmode
\epsfxsize=15cm
\epsfbox{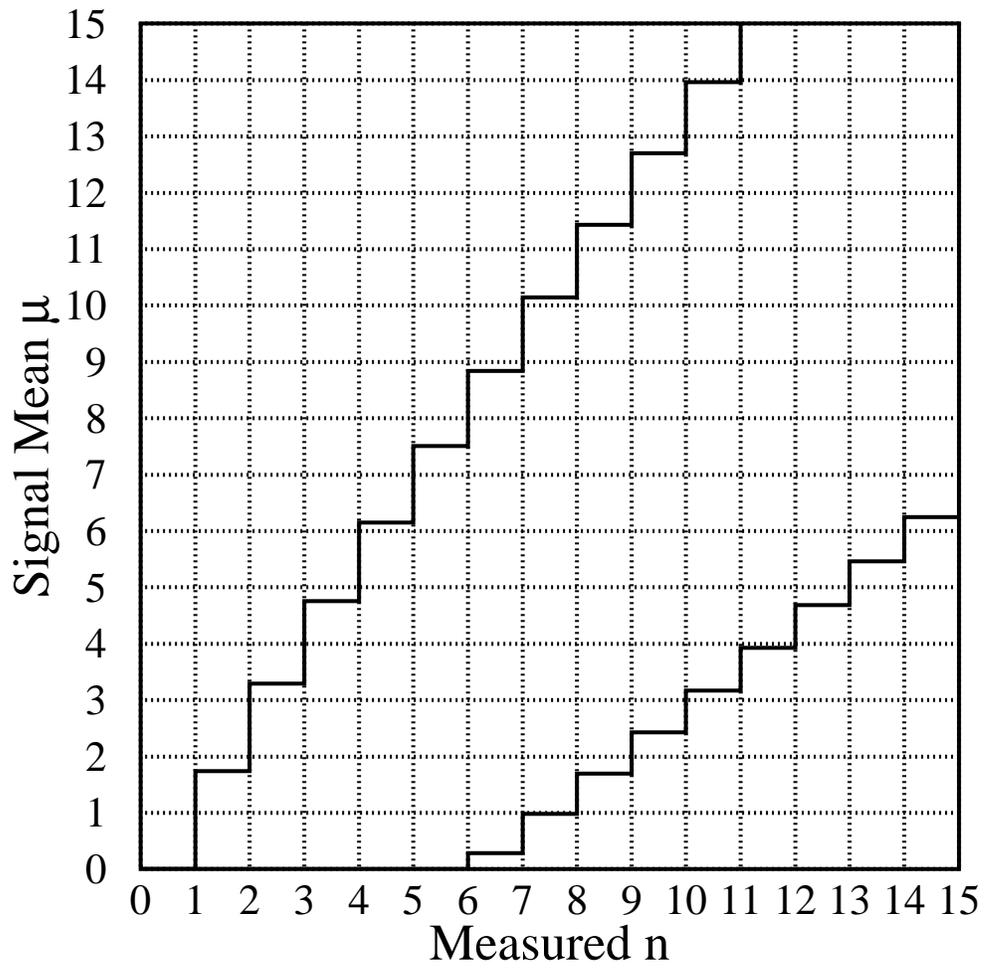}
\end{center}
\caption{Standard confidence belt for 90\% C.L. central confidence
intervals, for unknown Poisson signal mean $\mup$ in the presence
of Poisson background with known mean $\mub=3.0$.}
\label{fig-std-pois-central}
\end{figure}

Figures \ref{fig-std-pois-ul} and \ref{fig-std-pois-central} show
standard \cite{Garwood,Ricker} confidence belts for a Poisson process
when the observable $x$ is the total number of observed events $n$,
consisting of signal events with mean $\mup$ and
background events with {\em known} mean $\mub$. I.e.,

\begin{equation}
\label{eqn-pois}
P(n|\mup)  = (\mup+\mub)^n\exp(-(\mup+\mub))/n!
\end{equation}
In these figures, we use for illustration the case where 
$\mub=3.0$.

Since $n$ is an integer, Eq.\ (\ref{eqn-cover}) can only be
approximately satisfied.  By convention dating to the 1930's, one
strictly avoids undercoverage and replaces the equality in Eq.\
(\ref{eqn-cover}) with ``$\ge$''.  Thus the intervals overcover, and
are conservative.

Although the word ``conservative'' in this context may be viewed by
some as desirable, in fact it is an undesirable property of a set of
confidence intervals.  Ideal intervals cover the unknown true value
at exactly the stated confidence: 90\% C.L. intervals {\em should}
fail to contain the true value 10\% of the time.  If one desires
intervals which cover more than 90\% of the time, the solution is not
to add conservatism to the intervals, but rather to choose a higher
confidence level.  The discreteness of $n$ in the Poisson problem
leads unavoidably to some conservatism, but this is unfortunate, not a
virtue.

The Poisson intervals in Figs. \ref{fig-std-pois-ul} and
\ref{fig-std-pois-central} suffer from the same problems as the
Gaussian intervals.  First, if Physicist X uses the data to decide
whether to use Fig. \ref{fig-std-pois-ul} or
Fig. \ref{fig-std-pois-central}, then the resulting hybrid set can
undercover.  Second, there is a well-known problem if, for example,
$\mub=3.0$ and no events are observed.  In that case, the
confidence interval is again the empty set.  These problems are solved
by the ordering principle given in Sec.\ \ref{sec-order}.

For this Poisson case, there is an alternative set of intervals, given
by Crow and Gardner \cite{Crow}, which is instructive because it
requires the full Neyman construction.  In constructing these
intervals, one minimizes the horizontal length of the acceptance region
$[n_1,n_2]$ at each value of $\mup$.  Since $n$ is a discrete
variable, the concept of length in the horizontal direction can be
well-defined as the number of discrete points.  Said another way, the
points in the acceptance interval at each $\mup$ are chosen in
order of decreasing $P(n|\mup)$, until the until the sum of
$P(n|\mup)$ meets or exceeds the desired C.L.  (There are other
technical details in the original paper.)  The Crow-Gardner intervals
are instructive because neither Eq.\ (\ref{eqn-upper}) nor Eq.\
(\ref{eqn-central}) is satisfied, even as a conservative inequality.
(Recall that $x$ is identified with $n$ in this section.)  For $\alpha=0.9$,
$P(n<n_1|\mup)$ varies between 0.018 and 0.089, and
$P(n>n_2|\mup)$ varies between 0.011 and 0.078, in a manner
dictated by the Neyman construction so that always 
$P(n\in[n_1,n_2]\,|\mup) \ge 0.9$.  Like Crow and Gardner, we use 
Neyman's construction, but with a different ordering for choosing 
the points in the acceptance interval.

\section{New Intervals from an Ordering 
Principle Based on Likelihood Ratios}
\label{sec-order}
\subsection{Poisson with Background}

We begin with a numerical example which occurs in the construction of
confidence belts for a Poisson process with background.  The
construction proceeds in the manner of Fig.~\ref{fig-example-belt},
where the measurement $x$ in Fig.~\ref{fig-example-belt} now
corresponds to the measured total number of events $n$.

Let the known mean background be $\mub=3.0$, and consider the
construction of the horizontal acceptance interval at signal mean
$\mup=0.5$.  Then $P(n|\mup)$ is given by Eq.\ (\ref{eqn-pois}),
and is given in second column of Table~\ref{tab-pois-exam}.

Now consider, for example, $n=0$. For the assumed $\mub=3.$, the
probability of obtaining 0 events is 0.03 if $\mup=0.5$, which is
quite low on an absolute scale. However, it is not so low when
compared to the probability (0.05) of obtaining 0 events with
$\mub=3.$ and $\mup=0.0$, which is the alternate hypothesis with
the greatest likelihood.  A {\em ratio} of likelihoods, in this case
0.03/0.05, is what we use as our ordering principle when selecting
those values of $n$ to place in the acceptance interval.

That is, for each $n$, we let $\mubest$ be that value of mean signal
$\mup$ which maximizes $P(n|\mup)$; we require $\mubest$ to be
physically allowed, i.e., non-negative in this case.  Then $\mubest =
\max(0,n-\mub)$, and is given in the third column of
Table~\ref{tab-pois-exam}.  We then compute $P(n|\mubest)$, which is
given in the fourth column.  The fifth column contains the ratio,
\begin{equation}
\label{eqn-R-order}
R = P(n|\mup) / P(n|\mubest),
\end{equation}
and is the quantity on which our ordering principle is based.  $R$ is
a ratio of two likelihoods: the likelihood of obtaining $n$ given the
actual mean $\mup$, and the likelihood of obtaining $n$ given the best-fit
physically allowed mean.  Values of $n$ are added to the acceptance
region for a given $\mup$ in decreasing order of $R$, until the sum
of $P(n|\mup)$ meets or exceeds the desired C.L.  This ordering,
for values of $n$ necessary to obtain total probability of 90\%, is
shown in the column labeled ``rank''.  Thus, the acceptance region for
$\mup=0.5$ (analogous to a horizontal line segment in Figure 1), is
the interval $n=[0,6]$.  Due to the discreteness of $n$, the
acceptance region contains more summed probability than 90\%; this is
unavoidable no matter what the ordering principle, and leads to
confidence intervals which are conservative.

For comparison, in the column of Table~\ref{tab-pois-exam} labeled
``U.L.'', we place check marks at the values of $n$ which are in the
acceptance region of standard 90\% C.L. upper limits for this example;
and in the column labeled ``central'', we place check marks at the
values of $n$ which are in the acceptance region of standard 90\% C.L
central confidence intervals.

The construction proceeds by finding the acceptance region for all
values of $\mup$, for the given value of $\mub$.  With a
computer, we perform the construction on a grid of discrete values of
$\mup$, in the interval $[0,50]$ in steps of 0.005. This suffices
for the precision desired (0.01) in endpoints of confidence intervals.
We find that a mild pathology arises as a result of the fact that the
observable $n$ is discrete.  When the vertical dashed line is drawn
at some $n_0$ (in analogy with in Fig.~\ref{fig-example-belt}), it can
happen that the set of intersected horizontal line segments is not
simply connected.  When this occurs we naturally take the confidence
interval to have $\mu_1$ corresponding to the bottom-most segment
intersected, and to have $\mu_2$ corresponding to the top-most
segment intersected.

We then repeat the construction for a selection of fixed values of
$\mub$.  We find an additional mild pathology, again caused by the
discreteness in $n$: when we compare the results for different values
of $\mub$ for fixed $n_0$, the upper endpoint $\mu_2$ is not always
a decreasing function of $\mub$, as would be expected.  When this
happens, we force the function to be non-increasing, by lengthening
selected confidence intervals as necessary.  We have investigated this
behavior, and compensated for it, over a fine grid of $\mub$ in the
range $[0,25]$ in increments of 0.001 (with some additional searching
to even finer precision).

Our compensation for the two pathologies mentioned in the previous
paragraphs adds slightly to our intervals' conservatism, which however
remains dominated by the unavoidable effects due to the discreteness
in $n$.

\begin{figure}
\begin{center}
\leavevmode
\epsfxsize=15cm
\epsfbox{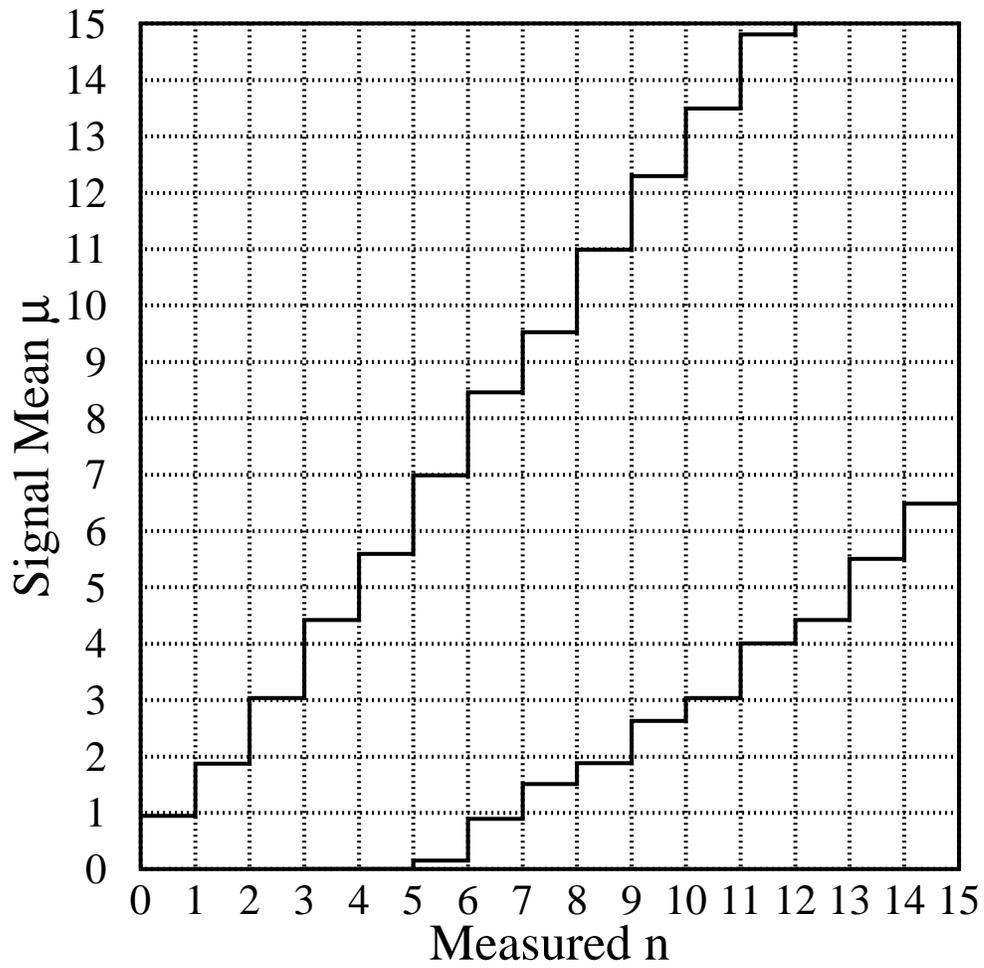}
\end{center}
\caption{Confidence belt based on our ordering principle, for 90\% C.L. 
confidence intervals for unknown Poisson signal mean $\mup$ in the
presence of Poisson background with known mean $\mub=3.0$.}
\label{fig-pois-new}
\end{figure}

The confidence belts resulting from our construction are shown in
Fig.~\ref{fig-pois-new}, which may be compared with Figs.
\ref{fig-std-pois-ul} and \ref{fig-std-pois-central}.
At large $n$, Fig.~\ref{fig-pois-new} is similar to
Fig.~\ref{fig-std-pois-central}; the background is effectively
subtracted without constraint, and our ordering principle produces
two-sided intervals which are approximately central intervals.  At
small $n$, the confidence intervals from Fig.~\ref{fig-pois-new}
automatically become upper limits on $\mup$; i.e., the lower endpoint
$\mu_1$ is 0 for $n\le4$ in this case.  Thus, flip-flopping between
Figs. \ref{fig-std-pois-ul} and \ref{fig-std-pois-central} is replaced
by one coherent set of confidence intervals, (and no interval is the
empty set).

Tables \ref{tab-p68a}-\ref{tab-p99b} give our confidence intervals
$[\mu_1,\mu_2]$ for the signal mean $\mup$ for the most commonly used
confidence levels, namely 68.27\% (sometimes called 1-$\sigma$
intervals by analogy with Gaussian intervals), 90\%, 95\%, and 99\%.
Values in italics indicate results which must be taken with particular
caution, since the probability of obtaining the number of events
observed or fewer is less than 1\%, even if $\mup=0$.  (See
Sec.~\ref{goodness-of-fit} below.)

Figure~\ref{fig-pdg-back1} shows, for $n=0$ through $n=10$, the value
of $\mu_2$ as a function of $\mub$, for 90\% C.L.  The small
horizontal sections in the curves are the result of the mild pathology
mentioned above, in which the original curves make a small dip, which
we have eliminated.  Dashed portions in the lower right indicate
results which must be taken with particular caution, corresponding to
the italicized values in the tables.  Dotted portions on the upper
left indicate regions where $\mu_1$ is non-zero.  These corresponding
values of $\mu_1$ are shown in Fig.~\ref{fig-pdg-back2}.

Figure~\ref{fig-pdg-back1} can be compared with the Bayesian
calculation in Fig.~28.8 of Ref.~\cite{PDG} which uses a uniform prior
for $\mutrue$.  A noticeable difference is that our curve for $n=0$
decreases as a function of $\mub$, while the result of the Bayesian
calculation stays constant (at 2.3).  The decreasing limit in our case
reflects the fact that $P(n_0|\mup)$ decreases as $\mub$
increases.  We find that objections to this behavior are typically
based on a misplaced Bayesian interpretation of classical intervals,
namely the attempt to interpret them as statements about
$P(\mutrue|n_0)$.

\subsection{Gaussian with Boundary at Origin}
\label{subsec-gauss-new}
It is straightforward to apply our ordering principle to the other
troublesome example of Sec.~\ref{sec-exam}, the case of a Gaussian
resolution function (Eq.~\ref{eqn-gauss}) for $\mup$, when
$\mup$ is physically bounded to non-negative values.  In analogy
with the Poisson case, for a particular $x$, we let $\mubest$ be the
physically allowed value of $\mup$ for which $P(x|\mup)$ is
maximum.  Then $\mubest = \max(0,x)$, and
\begin{equation}
\label{eqn-pmubest}
P(x|\mubest) = \left\{ \begin{array}{ll}
                       1/\sqrt{2\pi}, & \mbox{$x\ge0$}\\
                       \exp(-x^2/2)/\sqrt{2\pi}, & \mbox{$x<0$.}
                       \end{array}
               \right. 
\end{equation}
We then compute $R$ in analogy to Eq.~\ref{eqn-R-order}, using
Eqs.~\ref{eqn-gauss} and \ref{eqn-pmubest}:
\begin{equation}
\label{eqn-R-gauss}
R(x) = {P(x|\mup) \over P(x|\mubest) }
         = \left\{ \begin{array}{ll}
                   \exp(-(x-\mup)^2/2), & \mbox{$x\ge0$}\\
                   \exp(x\mup - \mup^2/2), & \mbox{$x<0$.}
                   \end{array}
           \right.
\end{equation}
During our Neyman construction of confidence intervals, $R$ determines
the order in which values of $x$ are added to the acceptance region at
a particular value of $\mup$.  In practice, this means that for a
given value of $\mup$, one finds the interval $[x_1,x_2]$ such that
$R(x_1)=R(x_2)$ and
\begin{equation}
\label{eqn-accept}
\int_{x_1}^{x_2} P(x|\mup) dx = \alpha.
\end{equation}
We solve for $x_1$ and $x_2$ numerically to the desired precision, for
each $\mup$ in a grid with 0.001 spacing.  With the acceptance regions
all constructed, we then read off the confidence intervals
$[\mu_1,\mu_2]$ for each $x_0$ as in Fig.~\ref{fig-example-belt}.

Table \ref{tab-gauss-new} contains the results for representative
measured values and confidence levels.  Figure~\ref{fig-gauss-new}
shows the confidence belt for 90\% C.L.

It is instructive to compare Fig.~\ref{fig-gauss-new} with
Fig.~\ref{fig-std-gauss-central}.  At large $x$, the confidence
intervals $[\mu_1,\mu_2]$ are the same in both plots, since that is
far away from the constraining boundary.  Below $x=1.28$, the lower
endpoint of the new confidence intervals is zero, so that there is
automatically a transition from two-sided confidence intervals to an
upper confidence limit given by $\mu_2$.  The point of this transition
is fixed by the calculation of the acceptance interval for
$\mup=0$; the solution has $x_1=-\infty$, and so
Eq.~\ref{eqn-accept} is satisfied by
$x_2=1.28$ when $\alpha=90\%$.  Of course, one is not obligated to
claim a non-null discovery just because the 90\% C.L. confidence
interval does not contain zero.  With a proper understanding of what
confidence intervals are (Sec.~\ref{sec-class}), one realizes that
they do not indicate degree of belief.

Our 90\% C.L. upper limit at $x=0$ is $\mu_2=1.64$, which,
interestingly, is the standard 95\% C.L. upper limit, rather than
$\mu_2=1.28$, which is the standard 90\% C.L. upper limit.  The
departure from the standard 90\% C.L. upper limits reflects the fact,
mentioned above, that they provide frequentist coverage only when the
decision to quote an upper limit is not based on the data.  Our method
repairs the undercoverage caused by flip-flopping
(Fig.\ref{fig-gauss-flipflop}), with a necessary cost in loosening the
upper limits around $x=0$.

As $x$ decreases, the upper limits from our method decrease,
asymptotically going as $1/|x|$ for large negative $x$.  As in the
Poisson case, particular caution is necessary when interpreting limits
obtained from measured values of $x$ which are unlikely for all
physical $\mup$.

\subsection{Decoupling of Goodness-of-Fit C.L. from the 
Confidence Interval C.L.}
\label{goodness-of-fit}

An advantage of our intervals compared to the standard classical
intervals is that ours effectively decouple the confidence level used
for a goodness-of-fit test from the confidence level used for
confidence interval construction.

To elaborate, let us first recall the procedure used in a standard
``easy'' $\chi^2$ fit (free from constraints, background, etc.), for
example the fit of a one-parameter curve to a set of points with
Gaussian error bars.  One examines the $\chi^2$ between the data and
the fitted curve, as function of the fit parameter.  The {\em value}
of $\chi^2$ at its minimum is used to determine goodness-of-fit: using
standard tables, one can convert this value to a goodness-of-fit
confidence level which tells one the quality of the fit.  A very poor
fit means that the information on the fitted parameter is suspect: the
experimental uncertainties may not be assessed properly, the
functional form of the parametrized curve may be wrong, or, in the
most general terms, the hypotheses being considered may not be the
relevant ones.

If the value of the minimum $\chi^2$ is considered acceptable, then
one examines the {\em shape} of $\chi^2$ (as a function of the fit
parameter) near its minimum, in order to obtain an (approximate)
confidence interval for the fit parameter at {\em any} desired
confidence level.  This procedure is powerful because it does not
permit random fluctuations that favor no particular parameter value to
influence the confidence interval.  The two confidence levels invoked
in this example are then independent; for example, one may require
that the goodness-of-fit C.L. be in the top 99\% in order to consider
the fit to be acceptable, while quoting a 68\% C.L.  confidence
interval for the fitted parameter.

The problems with the standard classical intervals in
Sec.~\ref{sec-exam} can be viewed from the point of view that they
effectively constrain the C.L. used for goodness-of-fit to be related
to that used for the confidence interval.  In both the Gaussian and
the Poisson upper limit examples, consider, for example, 90\% as the
C.L. for upper limits; the confidence interval is the empty set (or
outside the physical region, some prefer to say) some fraction of the
time which is determined by this choice of C.L.  
For example, if the true mean is zero in the
constrained Gaussian problem, then the empty set is obtained 10\% of
the time from Fig.~\ref{fig-std-gauss-ul}; if the true mean is zero in
the Poisson-with-background problem, the empty set can be obtained up
to 10\% of the time from confidence belts such as
Fig.~\ref{fig-std-pois-ul} (depending on the mean background $\mub$,
and on how discreteness affects the intervals for that $\mub$.)  An
empty-set confidence interval has the same effect as failing a
goodness-of-fit test: no useful confidence interval is inferred.  With
the standard confidence intervals, one is forced to use a specific
C.L. for this effective goodness-of-fit test, coupled to the C.L. used
for interval construction.  We believe this to be most undesirable,
and at the heart of the community's dissatisfaction with the standard
intervals.

In contrast, our construction always provides a confidence interval at
the desired confidence level (with of course some conservatism for the
discrete problems).  Independently, one can calculate the analog of
goodness-of-fit, and decide whether or not to consider the data or
model (including mean expected background) to be invalid.  This issue
arises in the case when an upper limit is quoted; i.e., the confidence
interval is $[0,\mu_2]$.  

In the constrained Gaussian case, one might
have data $x_0=-2.0$ and hence 90\% C.L. confidence interval $[0,0.4]$
from Tab.~\ref{tab-gauss-new}.  The natural analog for
goodness-of-fit is the probability to obtain $x\le x_0$ under the
best-fit assumption of $\mup=0$.  

In the Poisson-with-background
case, one might have data $n_0=1$ for $\mub=3$, and hence 90\%
C.L. confidence interval $[0,1.88]$ from Tab.~\ref{tab-p90a}.  The
natural analog for goodness-of-fit is the probability to obtain $n\le
n_0$ under the best-fit assumption of $\mup=0$.  

As noted above, in Fig.~\ref{fig-pdg-back1} we follow the practice of
the PDG \cite{PDG} by indicating with dashed lines those regions where
the goodness-of-fit criterion is less than
1\%.  In Tables \ref{tab-p68a}-\ref{tab-gauss-new}, the corresponding
intervals are italicized.

In summary, because our intervals decouple of the confidence level
used for a goodness-of-fit test from the confidence level used for
confidence interval construction, one is free to choose them
independently, at whatever level desired.

\section{Application to Neutrino Oscillation Searches}
\label{sec-neut}
\subsection{The Experimental Problem}
\label{sec-neut-prob}
Experimental searches for neutrino oscillations provide an example of
the application of this technique to a multidimensional problem.
Indeed it is just this problem that originally focused our attention
on this investigation.

Experiments of this type search for a transformation of one species of
neutrino into another.  To be concrete, we assume that the experiment
is to search for transformations between muon type neutrinos, $\nu_\mu
$, and electron type neutrinos, $\nu_e$, and that the influence of
other types of neutrinos can be ignored.  We hypothesize that the weak
eigenstates $\vert\nu_\mu\rangle$ and $\vert\nu_e\rangle$ are linear
superpositions of two mass eigenstates, $\vert\nu_1\rangle$ and
$\vert\nu_2\rangle$,
\begin{equation}
\vert\nu_e\rangle =
  \vert\nu_1\rangle \cos\theta+\vert\nu_2\rangle \sin\theta
\end{equation}
and
\begin{equation}
\vert\nu_\mu\rangle =
 \vert\nu_2\rangle \cos\theta-\vert\nu_1\rangle \sin\theta,
\end{equation}
and that the mass eigenvalues for $\vert\nu_1\rangle $ and
$\vert\nu_2\rangle $ are $m_1$ and $m_2$, respectively.  Quantum
mechanics dictates that the probability of such a transformation is
given by the formula \cite{PDG,Kayser}
\begin{equation}
\label{eqn-neut-osc-prob}
P(\nu_\mu\rightarrow\nu_e)=\sin^2(2\theta)\sin^2\left({{1.27\Delta
m^2L}\over E}\right),
\end{equation}
where $P$ is the probability for a $\nu_\mu$ to transform into a
$\nu_e$, $L$ is the distance in km between the creation of the
neutrino from meson decay and its interaction in the detector, $E$ is
the neutrino energy in GeV, and $\Delta m^2=|m_1^2 - m_2^2|$ in
$(\rm{eV}/c^2)^2$.

The result of such an experiment is typically represented as a
two-dimensional confidence region in the plane of the two unknown
physical parameters, $\theta$, the rotation angle between the weak and
mass eigenstates, and $\Delta m^2$, the (positive) difference between
the squares of the neutrino masses.  Traditionally, $\sin^2(2\theta)$
is plotted along the horizontal axis and $\Delta m^2$ is plotted along
the vertical axis.  An example of such a plot is shown in Fig.
\ref{fig-null-result}, based on a toy model that we develop below.  
In this example, no evidence for oscillations is seen and the
confidence region is set as the area to the left of the curve in this
figure.

\subsection{The Proposed Technique for Determining Confidence Regions}
\label{sec-neut-solution}
The problem of setting the confidence region for a neutrino
oscillation search experiment often shares all of the difficulties
discussed in the previous sections.  The variable $\sin^2(2\theta)$ is
clearly bounded by zero and one.  Values outside this region can have
no possible interpretation within the theoretical framework that
defines the unknown physical parameters.  Yet consider an experiment
searching in a region of $\Delta m^2$ in which oscillations either do
not exist or are well below the sensitivity of the experiment.  Such
an experiment is typically searching for a small signal of excess
$\nu_e$ interactions in a potentially large background of $\nu_e$
interactions from conventional sources and misidentified $\nu_\mu$
interactions.  Thus, it is equally likely to have a best fit to a
negative value of $\sin^2(2\theta)$ as to a positive one, provided
that the fit to Eq. (\ref{eqn-neut-osc-prob}) is unconstrained.

Typically, the experimental measurement consists of counting the
number of events in an arbitrary number of bins\cite{unbinned} in the
observed energy of the neutrino and possibly other measured variables,
such as the location of the interaction in the detector.  Thus, the
measured data consist of a set $N\equiv\{n_i\}$, together with an
assumed known mean expected background $B\equiv\{\mub_i\}$ and a
calculated expected oscillation contribution
$T\equiv\{\mup_i\vert \sin^2(2\theta), \Delta m^2\}$.

To construct the confidence region, the experimenter must choose an
ordering principle to decide which of the large number of possible $N$
sets should be included in the acceptance region for each point on the
$\sin^2(2\theta)-\Delta m^2$ plane.  We suggest the ordering principle
identical to the one suggested in Sec \ref{sec-order}, namely the
ratio of the probabilities
\begin{equation}
\label{eqn-neut-osc-R}
R={{P(N\vert T)}\over{P(N\vert \Tbest)}},
\end{equation}
where $\Tbest(\sin^2(2\theta)_{\rm best}, \Delta m^2_{\rm best})$
gives the highest probability for $P(N\vert T)$ for the physically
allowed values of $\sin^2(2\theta)$ and $\Delta m^2$.

In the Gaussian regime, $\chi^2=-2{\rm ln}(P)$, so this approach
is equivalent to using the difference in $\chi^2$ between $T$ and
$\Tbest$, i.e.,
\begin{equation}
\label{eqn-neut-osc-Rprime-one}
R^\prime\equiv\Delta
\chi^2=\sum_i\bigg[{{(n_i-\mub_i-\mup_i)^2}\over{\sigma_i^2}} -
{{(n_i-\mub_i-\mubesti)^2}\over{\sigma_i^2}} \bigg],
\end{equation}
where $\sigma_i$ is the Gaussian error.  We actually recommend an
alternative form based on the likelihood
function,\cite{Baker-and-Cousins}
\begin{equation}
\label{eqn-neut-osc-Rprime-two}
R^{\prime\prime}\equiv\Delta \chi^2=2\sum_i\bigg[\mup_i-\mubesti + n_i{\rm
ln}\bigg({{{\mubesti+\mub_i}\over{\mup_i+\mub_i}}}\bigg)\bigg],
\end{equation}
since it can be used in all cases.

To demonstrate how this works in practice, and how it compares to
alternative approaches that have been used, we consider a toy model of
a typical neutrino oscillation experiment.  The toy model is defined
by the following parameters: Mesons are assumed to decay to neutrinos
uniformly in a region 600~m to 1000~m from the detector.  The expected
background from conventional $\nu_e$ interactions and misidentified
$\nu_\mu$ interactions is assumed to be 100 events in each of 5 energy
bins which span the region from 10 to 60 GeV.  We assume that the
$\nu_\mu$ flux is such that if $P(\nu_\mu\rightarrow\nu_e) = 0.01$
averaged over any bin, then that bin would have an expected additional
contribution of 100 events due to $\nu_\mu\rightarrow\nu_e$
oscillations.

The acceptance region for each point in the $\sin^2(2\theta)-\Delta
m^2$ plane is calculated by performing a Monte Carlo simulation of
the results of a large number of experiments for the given set of
unknown physical parameters and the known neutrino flux of the actual
experiment.  For each experiment, $\Delta \chi^2$ is calculated
according to the prescription of either
Eq.~\ref{eqn-neut-osc-Rprime-one} or \ref{eqn-neut-osc-Rprime-two}.
The single number that is needed for each point in the
$\sin^2(2\theta)-\Delta m^2$ plane is $\Delta
\chi^2_c(\sin^2(2\theta),\Delta m^2)$, such that $\alpha$ of the
simulated experiments have $\Delta \chi^2<\Delta \chi^2_c$.  After the
data are analyzed, $\Delta \chi^2$ for the data and each point in the
$\sin^2(2\theta)-\Delta m^2$ plane, i.e. $\Delta
\chi^2(N|\sin^2(2\theta),\Delta m^2)$,  is 
compared to $\Delta \chi^2_c$ and the acceptance region is all points
such that
\begin{equation}
\Delta \chi^2(N|\sin^2(2\theta),\Delta m^2)<\Delta 
\chi^2_c(\sin^2(2\theta),\Delta m^2). 
\end {equation}

Figure \ref{fig-null-result} is an example of the result of a
calculation for a random experiment in the toy model for which there
were no oscillations, i.e., for $\sin^2(2\theta) = 0$.

One might naively expect that $\Delta \chi^2_c = 4.61$, the 90\%
C.L. value for a $\chi^2$ distribution with two degrees of freedom.
For the toy model, it actually varies from about 2.4 to 6.6 across the
$\sin^2(2\theta)-\Delta m^2$ plane.  The deviation from 4.61 is caused
by at least three effects:
\begin{enumerate}
\item
Proximity to the unphysical region.  Points close to the unphysical
region occasionally have best fits in the unphysical region.  Since
our algorithm restricts fits to the physical region, these fits give a
lower $\Delta\chi^2$ than unrestricted fits.
\item
Sinusoidal nature of the oscillation function.  The $\chi^2$
distribution assumes a Gaussian probability density function,
but the oscillation probability function is sinusoidal.  For high
values of $\Delta m^2$ fluctuations can cause a global minimum in a
``wrong" trough of the function, increasing the value of $\Delta
\chi^2$ from what it would be if there were only one trough.
\item
One-dimensional regions.  In some regions of the plane, the
probability distribution function becomes one rather than two
dimensional.  For example, at very low values of $\Delta m^2$ the only
relevant quantity is the number of events in the lowest energy bin,
since the oscillation probability, Eq.~( \ref{eqn-neut-osc-prob}), is
proportional to $1/E^2$ for sufficiently low $\Delta m^2$.
Fluctuations in higher energy bins do not lead to any physical
interpretation, and thus cancel in the calculation of $\Delta \chi^2$.
In these regions, $\Delta \chi^2_c$ tends to lower values than normal.
\end{enumerate}

\subsection{Comparison to Alternative Classical Methods}
\label{sec-neut-compare}
Most papers reporting the results of neutrino oscillation searches
have not been explicit enough for us to determine exactly how the
confidence regions were set.  However, we can imagine three classical
methods that either have or could have been used.  We refer to these
as the Raster Scan, the Flip-Flop Raster Scan, and the Global Scan.
All of them have the advantage that a Gaussian approximation is made
so that a full Neyman construction of the confidence region is not
necessary.
\begin{enumerate}
\item
The Raster Scan: For each value of $\Delta m^2$, a best fit is made
for $\sin^2(2\theta)$. At each $\Delta m^2$, $\chi^2$ is calculated as
a function of $\sin^2(2\theta)$, and the 1-D confidence interval in
$\sin^2(2\theta)$ at that $\Delta m^2$ is taken to be all points that
have a $\chi^2$ within 2.71 of the minimum value. (2.71 is the
two-sided 90\% C.L.\ for a $\chi^2$ distribution with one degree of
freedom.)  The confidence region in the $(\sin^2(2\theta),\Delta m^2)$
plane is then the union of all these intervals.
\item
The Flip-Flop Raster Scan: Similar to the Raster Scan except that a
decision to use a one-sided upper limit or a two-sided interval is
made based on the data.  If there is a signal with significance
greater than three standard deviations, the Raster Scan is used.  If
not, an upper limit is set by a raster scan using the one-sided 90\%
C.L.\ $\Delta \chi^2$ value of 1.64.
\item
The Global Scan: A best fit is made to both $\sin^2(2\theta)$ and
$\Delta m^2$, and the confidence region is given as all points that
have a $\chi^2$ within 4.61 of the minimum value. (As mentioned above,
4.61 is the two-sided 90\% C.L.\ for a $\chi^2$ distribution with two
degrees of freedom.)
\end{enumerate}

In all three cases, we assume that there is no restriction that the
best fit be in the physical region.  This is because the method of
using a fixed $\Delta\chi^2$ depends on the reference $\chi^2$ being
the minimum of a parabolic $\chi^2$ distribution.  Any attempt to
restrict the minimum to the physical region automatically gives
improper coverage.  Thus, all three of these methods suffer from the
possibility that they could either rule out the entire physical plane,
or give limits which are not characteristic of the sensitivity of the
experiment.

We have used the toy model to study the coverage of each of these
techniques.  The Raster Scan gives exact coverage.  However, it is not
a powerful technique in that it cannot distinguish a likely value of
$\Delta m^2$ from an unlikely one, since it works at fixed $\Delta
m^2$.  This is best illustrated in the case in which a positive signal
is found.  Figure \ref{fig-power-demo} shows the confidence regions
for both the Raster Scan and our proposed technique for a sample case
for which $\Delta m_t^2=40$ (eV/$c^2)^2$ and $\sin^2(2\theta_t)=0.006$.
Both techniques provide exact coverage, but the proposed technique
isolates the signal, with one ghost region, while the Raster Scan does
not.

Since the Raster Scan gives exact coverage, it will not surprise the
reader to learn that the Flip-Flop Raster Scan undercovers for the
reasons given in Sec.~\ref{sec-exam}.  Figure
\ref{fig-flipflop-cover} shows the region of significant
undercoverage (greater than 1\%) for the Flip-Flop Raster Scan.  The
coverage drops as low as 85\%, as is to be expected from the
discussion in Sec.~\ref{sec-exam}.  To set the scale, a quantity we
call the ``sensitivity" is also shown in this figure.   The sensitivity
is defined as the average upper limit one would get from an ensemble 
of experiments with the expected background and no true signal.
We discuss the use of this
quantity further in Sec.~\ref{sec-excess}.

Unlike the Raster Scan and Flip-Flop Raster Scan, the Global Scan is a
powerful technique.  However, it suffers from not giving proper
coverage for the reasons enumerated at the end of the previous
subsection (numbers 2 and 3).  It has both regions of undercoverage
and overcoverage, as shown in Fig.~\ref{fig-global-cover}.  The
coverage varies across the plane from about 76\% to 94\%.

Table \ref{tab-Brand-X} summarizes the properties of the proposed
technique and the three alternative techniques that we have
considered.

\section{The Problem of Fewer Events than Expected Background}
\label{sec-excess}
We started this investigation to solve the problem in classical
statistics in which an experiment which measures significantly fewer
events than are expected from backgrounds will report a meaningless or
unphysical result.  While we have solved that problem, our solution
still yields results that are bothersome to some in that an experiment
that measures fewer events than expected from backgrounds will report
a lower upper limit than an identical experiment that measures a
number of events equal to that expected from background.  This seems
particularly troublesome in the case in which the experiment has no
observed events.  Why should an experiment claim credit for expected
backgrounds, when it is clear, in that particular experiment, there
were none?  Or why should a well designed experiment which has no
background and observes no events be forced to report a higher upper
limit than a less well designed experiment which expects backgrounds,
but, by chance, observes none?

The origin of these concerns lies in the natural tendency to want to
interpret these results as the probability $P(\mutrue|x_0)$ of a
hypothesis given data, rather than what they are really related to,
namely the probability $P(x_0|\mu)$ of obtaining data given a
hypothesis.  It is the former that a scientist may want to know in
order to make a decision, but the latter which classical confidence
intervals relate to.  As we discussed in Sec.~\ref{sec-bayes},
scientists may make Bayesian inferences of $P(\mutrue|x_0)$ based on
experimental results combined with their personal, subjective prior
probability distribution function.  It is thus incumbent on the
experimenter to provide information that will assist in this
assessment.

Our suggestion for doing this is that in cases in which the
measurement is less than the estimated background, the experiment
report both the upper limit and the ``sensitivity'' of the experiment,
where the ``sensitivity'' is defined as the average upper limit that
would be obtained by an ensemble of experiments with the expected
background and no true signal.  Table \ref{tab-sens} gives these
values, for the case of a measurement of a Poisson variable.

Thus, an experiment that measures 2 events and has an expected
background of 3.5 events would report a 90\% C.L. upper limit of 2.7
events (from Tab.~\ref{tab-p90a}), but a sensitivity of 4.6 events
(from Tab.~\ref{tab-sens}).

Figure \ref{fig-null-result-sens} represents a common occurrence for a
neutrino oscillation search experiment.  It is a repeat of
Fig.~\ref{fig-null-result}, an example of the toy model in which
$\sin^2(2\theta)=0$, but with the sensitivity shown by a dashed line.
The behavior is typical of what one would expect.  Due to random
fluctuations, the upper limit is greater than the sensitivity for some
values of $\Delta m^2$ and less than others.  In this case, it is due
to fluctuations, but in an actual experiment, it could also be due to
the presence of a signal around or below the experiment's sensitivity
at some value $\Delta m^2$, making other values of $\Delta m^2$ less
likely.  Again, for cases in which a significant portion of the upper
limit curve is below the sensitivity of the experiment, we suggest
that the sensitivity curve be displayed as well as the upper limit.

\section{Conclusion}
\label{sec-conclude}
The construction described in this paper strictly adheres to the Neyman
method \cite{Neyman}, as applied to discrete distributions since the
1930's \cite{Garwood,Ricker,Crow}.  Thus, the resulting confidence
intervals are firmly grounded in classical statistics theory.  What is
new is the particular choice of ordering we make within the freedom
inherent in Neyman's method.  This choice, described in
Sec.~\ref{sec-order}, yields intervals which automatically change over
from upper limits to two-sided intervals as the ``signal'' becomes
more statistically significant.  This eliminates
undercoverage caused by basing this choice on the data
(``flip-flopping'').  Our tables give classical confidence intervals
for the two common problems for which the PDG has described Bayesian
solutions incorporating a (questionable) uniform prior for a bounded
variable: Poisson processes with background, and Gaussian errors with
a bounded physical region.  This introduction of Bayesian methods was
at least partly motivated by problems with the traditional classical
intervals (non-physical or empty-set intervals, and coupling of
goodness-of-fit C.L. with confidence interval C.L.) which our new
intervals solve.  Thus, there should be renewed discussion of the
appropiateness of Bayesian intervals for reporting experimental
measurements in an objective way.

The new ordering principle can be applied quite generally. We have
developed the application to neutrino oscillation searches, where the
confidence region can have a particularly complicated stucture due to
physical constraints and multiple local minima in the pdf's.

Finally, we certainly agree that no matter how one constructs an
interval, it is important to publish relevant ingredients to the
calculation so that the reader (and the PDG) can (at least
approximately) perform alternative calculations or combine the result
with other experiments \cite{James-neut}.  In the Gaussian case, the
ingredients are the measured value (even if non-physical) and the
standard error. (Separating the statistical and systematic errors, as
is often done, is even better).  In the case of a counting experiment
with known background, the required ingredients are: the number of
observed events; the expected mean background; and the factor
(incorporating, e.g., integrated luminosity, efficiencies, etc.,)
which converts the number of observed events to the relevant physics
quantity (cross section, branching ratio, etc.).

{\em Note added in proof.}
Although we are not aware of previous application of this ordering
principle to the construction of confidence intervals for the
presentation of scientific results, such an ordering is naturally
implied by the theory of likelihood ratio tests, as explained in
Sec. 23.1 of Ref. [10].  We thank H. Chernoff for clarifying 
discussions on this point.

\acknowledgments

We thank Frederick James, Frederick Weber, and Sanjib Mishra for
useful discussions and comments on the manuscript.  This work was
supported by the U.S. Department of Energy.

\clearpage

\begin{figure}
\begin{center}
\leavevmode
\epsfxsize=15cm
\epsfbox{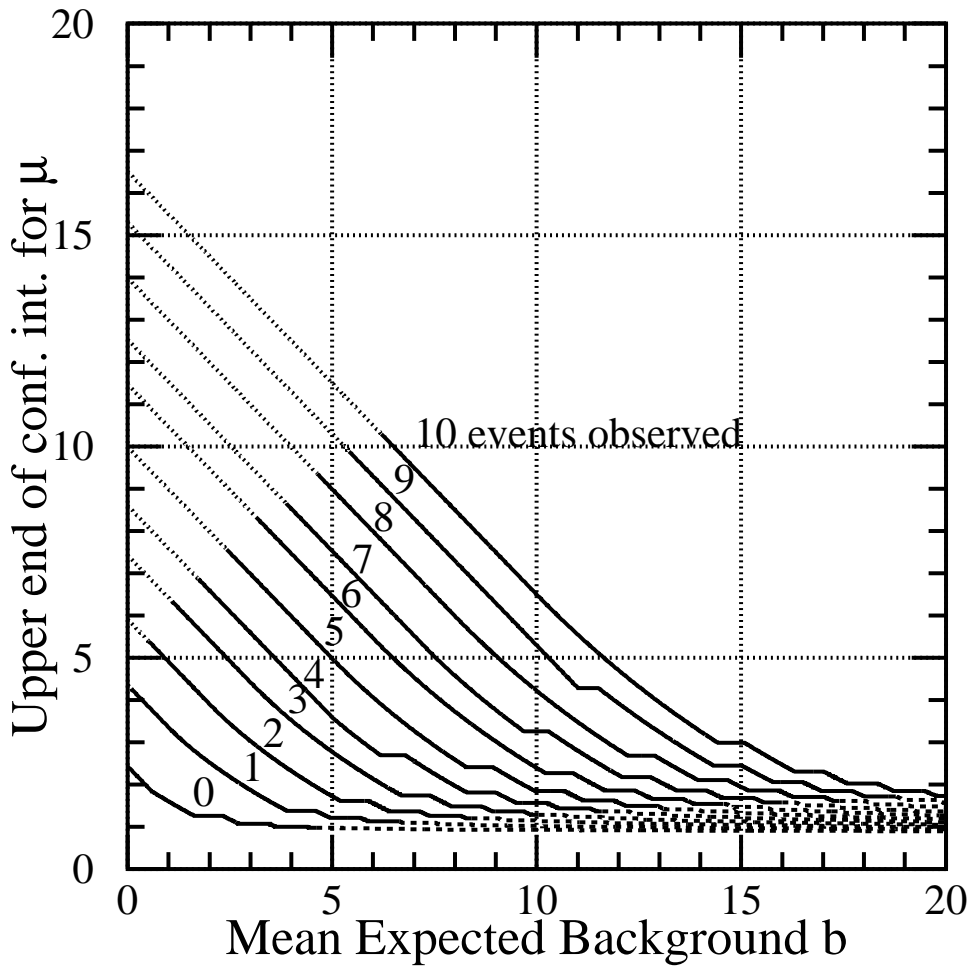}
\end{center}
\caption{Upper end $\mu_2$ of our 90\% C.L. confidence 
intervals $[\mu_1,\mu_2]$, for unknown Poisson signal mean $\mup$
in the presence of expected Poisson background with known mean
$\mub$.  The curves for the cases $n_0$ from 0 through 10 are
plotted.  Dotted portions on the upper left indicate regions where
$\mu_1$ is non-zero (and shown in the following figure).  Dashed
portions in the lower right indicate regions where the probability of
obtaining the number of events observed or fewer is less than 1\%,
even if $\mup=0$.  }
\label{fig-pdg-back1}
\end{figure}

\begin{figure}
\begin{center}
\leavevmode
\epsfxsize=15cm
\epsfbox{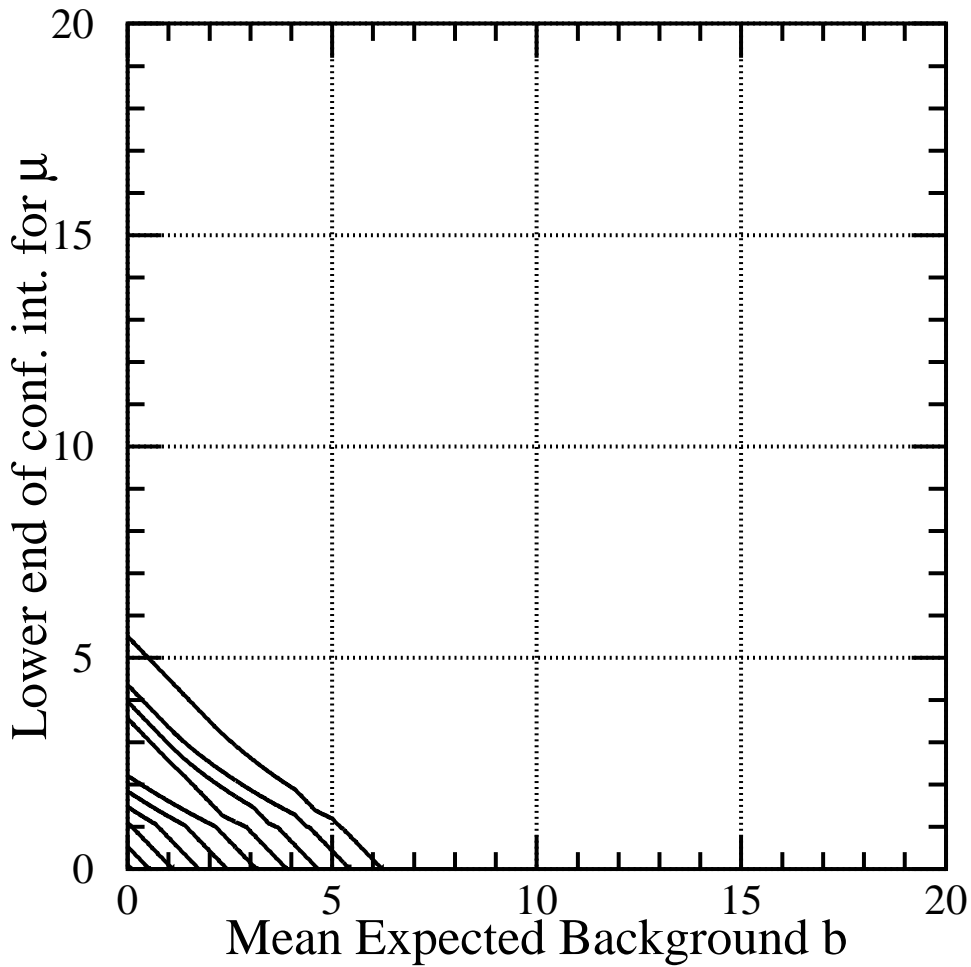}
\end{center}
\caption{Lower end $\mu_1$ of our 90\% C.L. confidence
intervals $[\mu_1,\mu_2]$, for unknown Poisson signal mean $\mup$
in the presence of expected Poisson background with known mean
$\mub$.  The curves correspond to the dotted regions in the plots
of $\mu_2$ of the previous figure, with again $n_0=10$ for the upper
right curve, etc.}
\label{fig-pdg-back2}
\end{figure}

\begin{figure}
\begin{center}
\leavevmode
\epsfxsize=15cm
\epsfbox{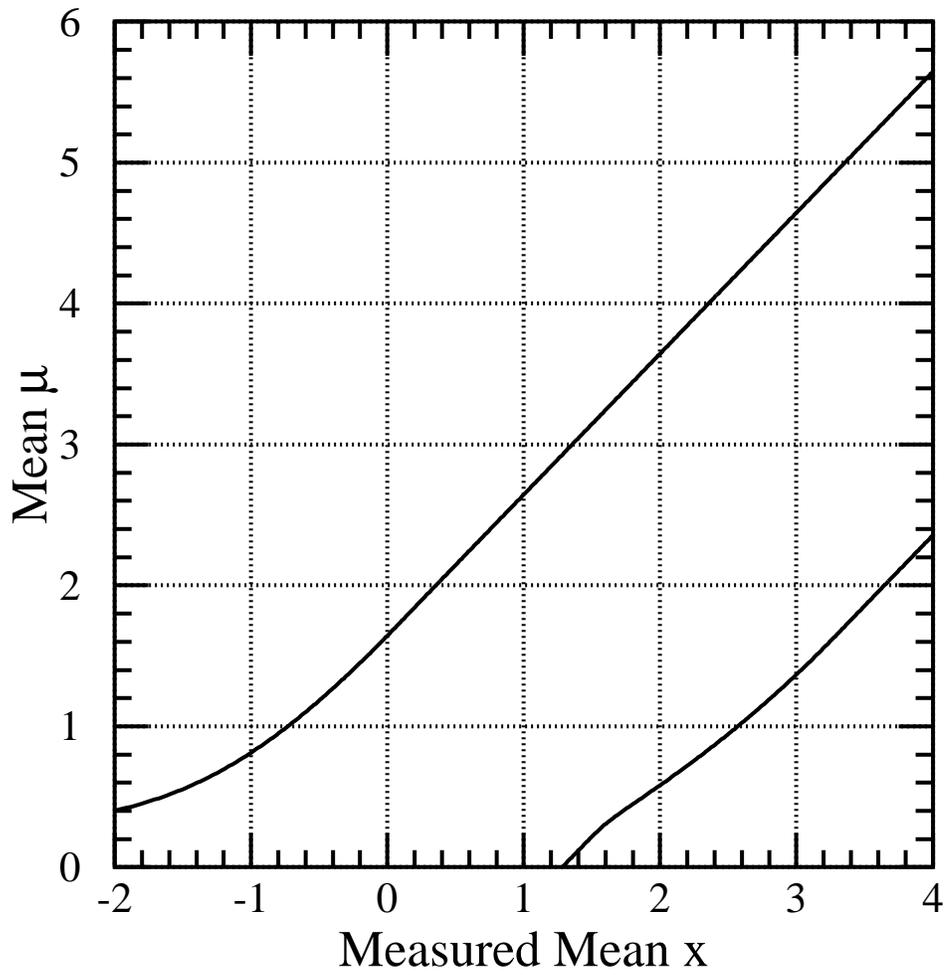}
\end{center}
\caption{Plot of our 90\% confidence intervals for mean of a Gaussian,
constrained to be non-negative, described in the text.  }
\label{fig-gauss-new}
\end{figure}

\begin{figure}
\begin{center}
\leavevmode
\epsfxsize=15cm
\epsfbox{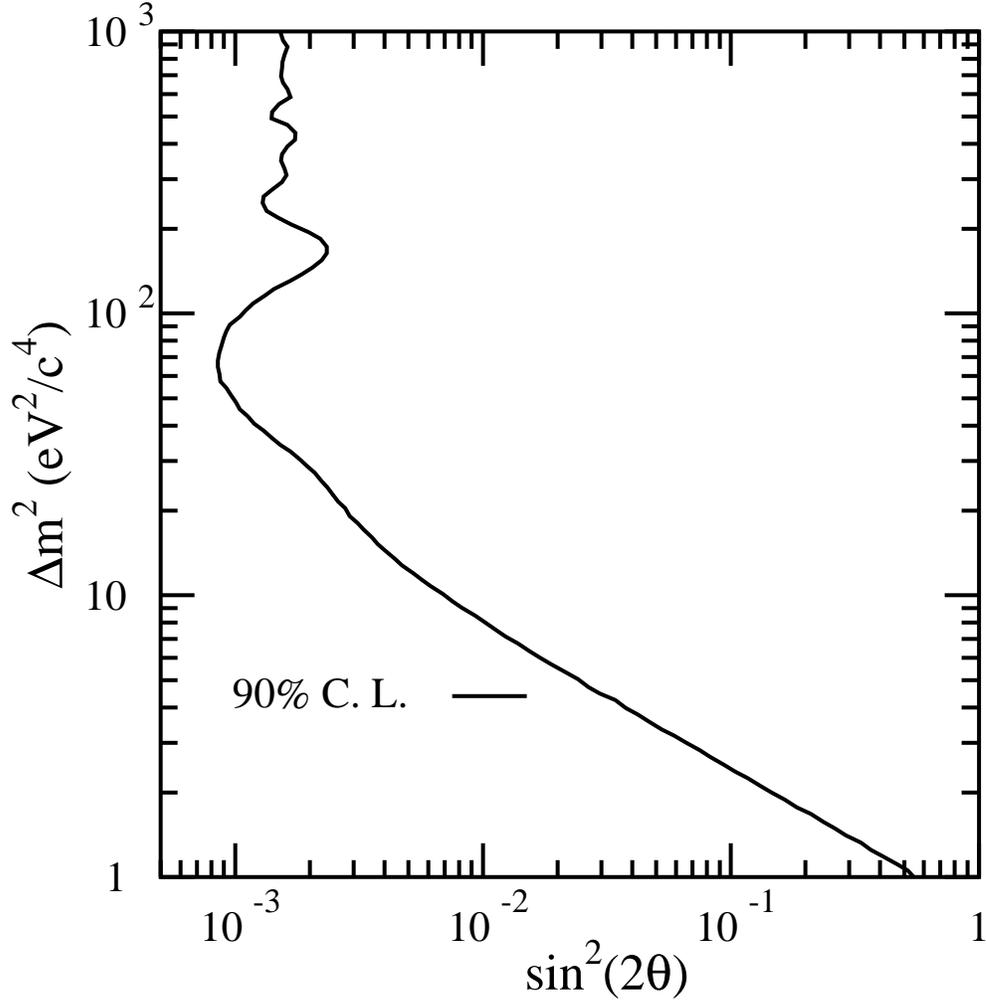}
\end{center}
\caption{Calculation of the confidence region for an example of the toy model 
in which $\sin^2(2\theta)=0$.  The 90\% confidence region is the area
to the left of the curve. }
\label{fig-null-result}
\end{figure}

\begin{figure}
\begin{center}
\leavevmode
\epsfxsize=15cm
\epsfbox{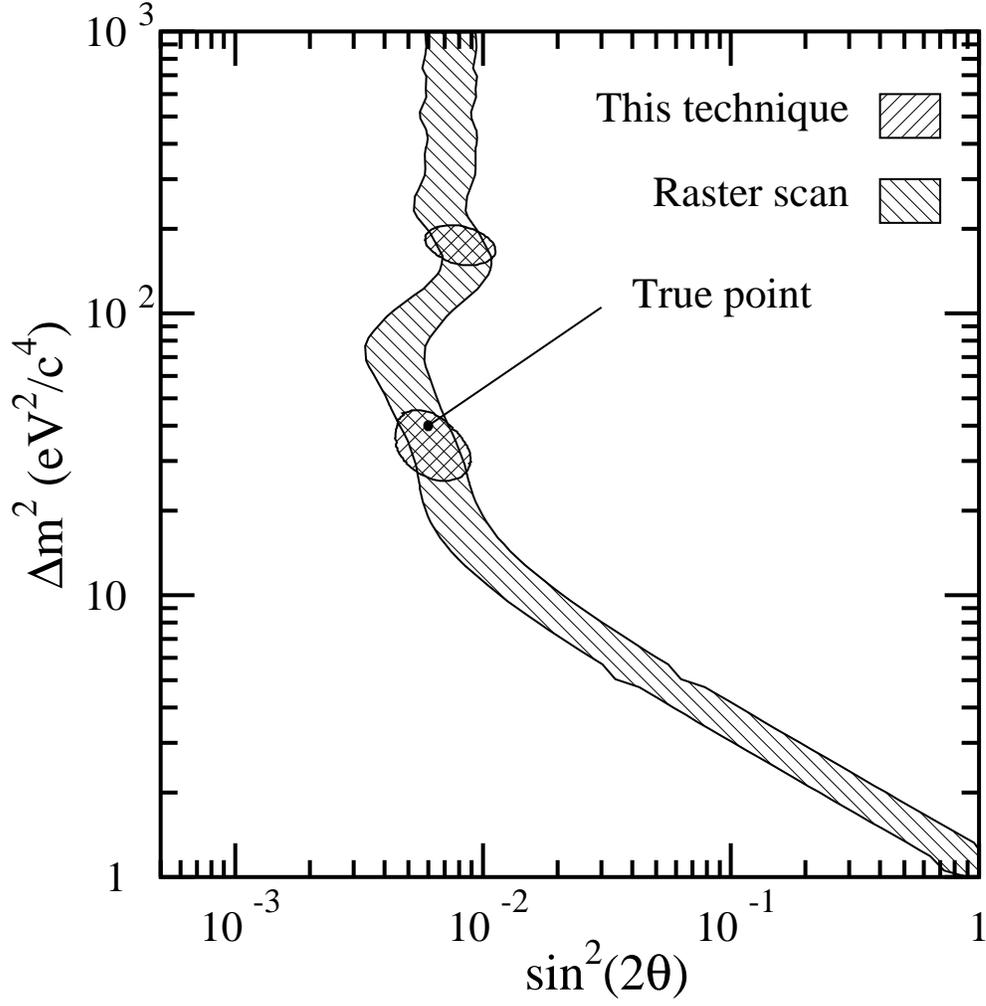}
\end{center}
\caption{Calculation of the confidence regions for an example of the toy model 
in which $\Delta m^2=40$ (eV/$c^2)^2$ and $\sin^2(2\theta)=0.006$, as
evaluated by the proposed technique and the Raster Scan.  }
\label{fig-power-demo}
\end{figure}

\begin{figure}
\begin{center}
\leavevmode
\epsfxsize=15cm
\epsfbox{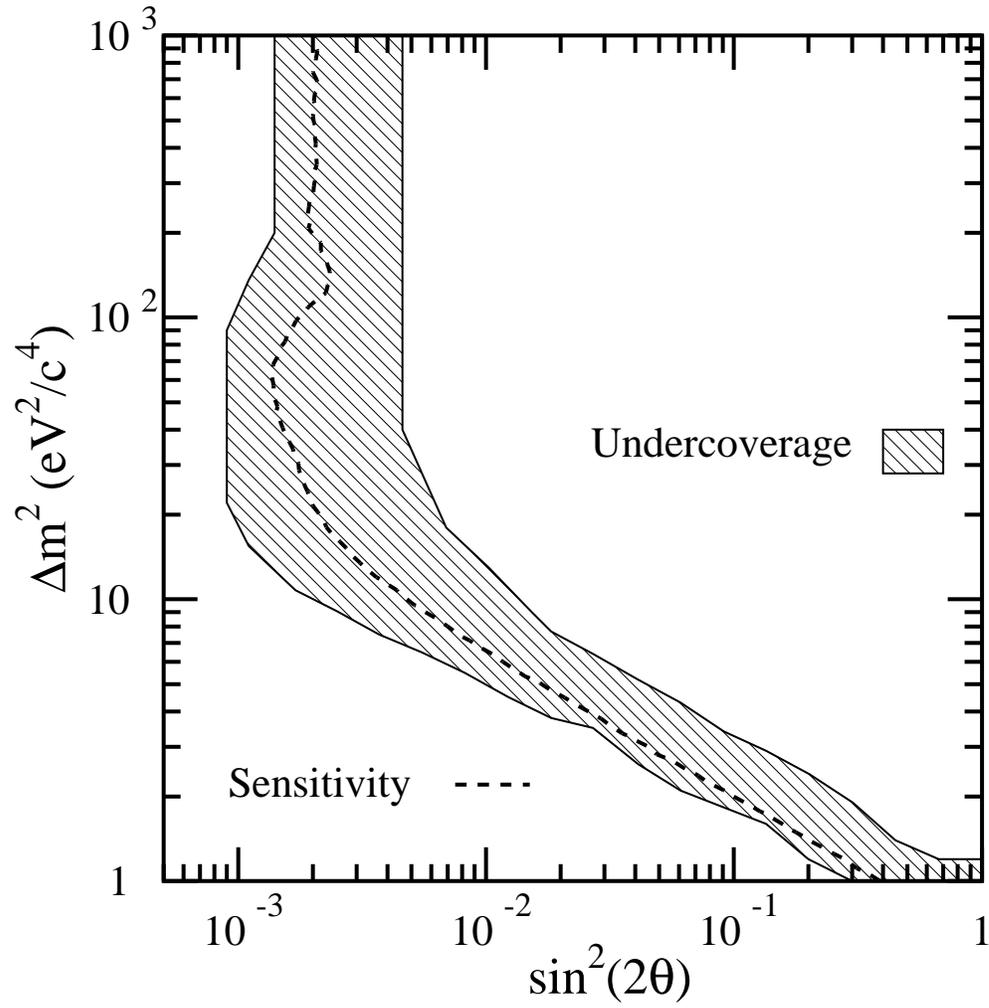}
\end{center}
\caption{Region of significant undercoverage for the Flip-Flop 
Raster Scan.}
\label{fig-flipflop-cover}
\end{figure}

\begin{figure}
\begin{center}
\leavevmode
\epsfxsize=15cm
\epsfbox{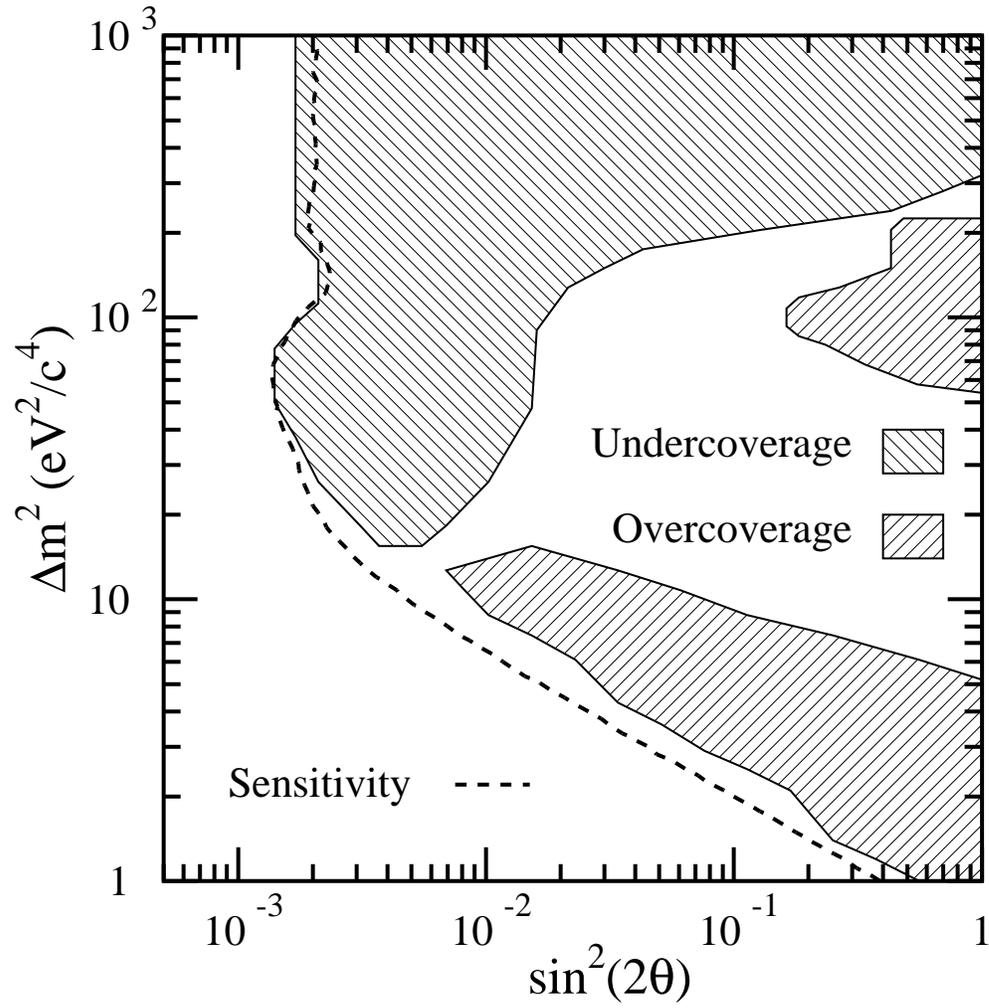}
\end{center}
\caption{Regions of significant under- and overcoverage for the 
Global Scan.}
\label{fig-global-cover}
\end{figure}

\begin{figure}
\begin{center}
\leavevmode
\epsfxsize=15cm
\epsfbox{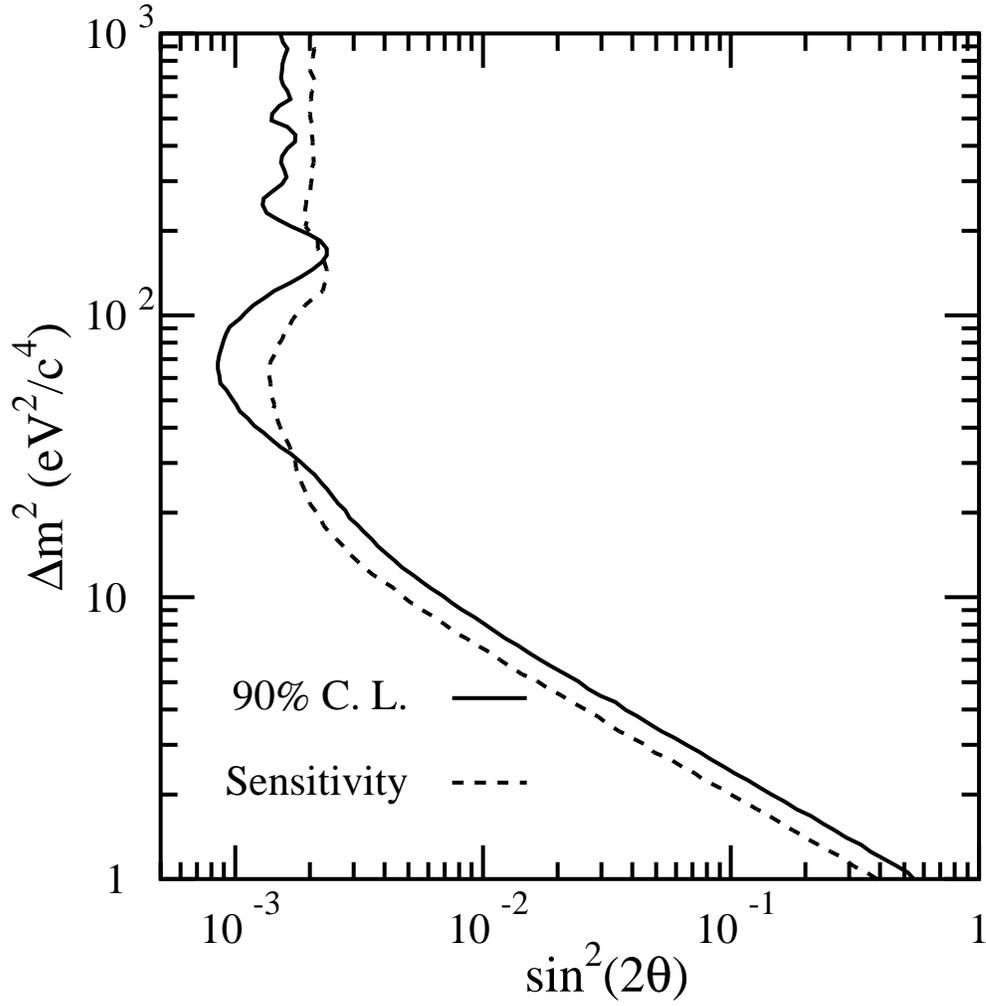}
\end{center}
\caption{Comparision of the confidence region for an example of 
the toy model in which $\sin^2(2\theta)=0$ and the sensitivity of the
experiment, as defined in the text.}
\label{fig-null-result-sens}
\end{figure}

\clearpage
\begin{table}
\caption{Illustrative calculations in the confidence belt 
construction for signal mean $\mup$ in the presence of known
mean background $\mub=3.0$.  Here we find the acceptance interval
for $\mup=0.5$.}
\label{tab-pois-exam}
\bigskip
\begin{tabular}{cccccccc} \hline
~$n$~ & $P(n|\mup)$ & $\mubest$ & $P(n|\mubest)$ & $R$ 
& rank & U.L. & central \\ \hline
 0 & 0.030 & 0. & 0.050 & 0.607 & 6 &         &         \\ 
 1 & 0.106 & 0. & 0.149 & 0.708 & 5 & $\surd$ & $\surd$ \\ 
 2 & 0.185 & 0. & 0.224 & 0.826 & 3 & $\surd$ & $\surd$ \\ 
 3 & 0.216 & 0. & 0.224 & 0.963 & 2 & $\surd$ & $\surd$ \\ 
 4 & 0.189 & 1. & 0.195 & 0.966 & 1 & $\surd$ & $\surd$ \\ 
 5 & 0.132 & 2. & 0.175 & 0.753 & 4 & $\surd$ & $\surd$ \\
 6 & 0.077 & 3. & 0.161 & 0.480 & 7 & $\surd$ & $\surd$ \\ 
 7 & 0.039 & 4. & 0.149 & 0.259 &   & $\surd$ & $\surd$ \\ 
 8 & 0.017 & 5. & 0.140 & 0.121 &   & $\surd$ &   	   \\ 
 9 & 0.007 & 6. & 0.132 & 0.050 &   & $\surd$ &   	   \\
10 & 0.002 & 7. & 0.125 & 0.018 &   & $\surd$ &   	   \\ 
11 & 0.001 & 8. & 0.119 & 0.006 &   & $\surd$ &   	    
\end{tabular}
\end{table}

\clearpage

{\squeezetable
\widetext
\begin{table}
\caption{Our 68.27\% C.L. intervals for the Poisson signal mean 
$\mup$, for total events observed $n_0$, for known mean background
$\mub$ ranging from 0 to 5.}
\label{tab-p68a}
\bigskip
\begin{tabular}{r|cccccccccc} \hline
~$n_0\backslash\mub$~  &    0.0 &    0.5 &    1.0 &    1.5 &    2.0 
&    2.5 &    3.0 &    3.5 &    4.0 &    5.0\\ \hline
  0 &  0.00, 1.29 &  0.00, 0.80 &  0.00, 0.54 &  0.00, 0.41 &  0.00, 0.41 
&  0.00, 0.25 &  0.00, 0.25 &  0.00, 0.21 &  0.00, 0.21 & {\it 0.00, 0.19} \\
  1 &  0.37, 2.75 &  0.00, 2.25 &  0.00, 1.75 &  0.00, 1.32 &  0.00, 0.97 
&  0.00, 0.68 &  0.00, 0.50 &  0.00, 0.50 &  0.00, 0.36 &  0.00, 0.30 \\
  2 &  0.74, 4.25 &  0.44, 3.75 &  0.14, 3.25 &  0.00, 2.75 &  0.00, 2.25 
&  0.00, 1.80 &  0.00, 1.41 &  0.00, 1.09 &  0.00, 0.81 &  0.00, 0.47 \\
  3 &  1.10, 5.30 &  0.80, 4.80 &  0.54, 4.30 &  0.32, 3.80 &  0.00, 3.30 
&  0.00, 2.80 &  0.00, 2.30 &  0.00, 1.84 &  0.00, 1.45 &  0.00, 0.91 \\
  4 &  2.34, 6.78 &  1.84, 6.28 &  1.34, 5.78 &  0.91, 5.28 &  0.44, 4.78 
&  0.25, 4.28 &  0.00, 3.78 &  0.00, 3.28 &  0.00, 2.78 &  0.00, 1.90 \\
  5 &  2.75, 7.81 &  2.25, 7.31 &  1.75, 6.81 &  1.32, 6.31 &  0.97, 5.81 
&  0.68, 5.31 &  0.45, 4.81 &  0.20, 4.31 &  0.00, 3.81 &  0.00, 2.81 \\
  6 &  3.82, 9.28 &  3.32, 8.78 &  2.82, 8.28 &  2.32, 7.78 &  1.82, 7.28 
&  1.37, 6.78 &  1.01, 6.28 &  0.62, 5.78 &  0.36, 5.28 &  0.00, 4.28 \\
  7 &  4.25,10.30 &  3.75, 9.80 &  3.25, 9.30 &  2.75, 8.80 &  2.25, 8.30 
&  1.80, 7.80 &  1.41, 7.30 &  1.09, 6.80 &  0.81, 6.30 &  0.32, 5.30 \\
  8 &  5.30,11.32 &  4.80,10.82 &  4.30,10.32 &  3.80, 9.82 &  3.30, 9.32 
&  2.80, 8.82 &  2.30, 8.32 &  1.84, 7.82 &  1.45, 7.32 &  0.82, 6.32 \\
  9 &  6.33,12.79 &  5.83,12.29 &  5.33,11.79 &  4.83,11.29 &  4.33,10.79 
&  3.83,10.29 &  3.33, 9.79 &  2.83, 9.29 &  2.33, 8.79 &  1.44, 7.79 \\
 10 &  6.78,13.81 &  6.28,13.31 &  5.78,12.81 &  5.28,12.31 &  4.78,11.81 
&  4.28,11.31 &  3.78,10.81 &  3.28,10.31 &  2.78, 9.81 &  1.90, 8.81 \\
 11 &  7.81,14.82 &  7.31,14.32 &  6.81,13.82 &  6.31,13.32 &  5.81,12.82 
&  5.31,12.32 &  4.81,11.82 &  4.31,11.32 &  3.81,10.82 &  2.81, 9.82 \\
 12 &  8.83,16.29 &  8.33,15.79 &  7.83,15.29 &  7.33,14.79 &  6.83,14.29 
&  6.33,13.79 &  5.83,13.29 &  5.33,12.79 &  4.83,12.29 &  3.83,11.29 \\
 13 &  9.28,17.30 &  8.78,16.80 &  8.28,16.30 &  7.78,15.80 &  7.28,15.30 
&  6.78,14.80 &  6.28,14.30 &  5.78,13.80 &  5.28,13.30 &  4.28,12.30 \\
 14 & 10.30,18.32 &  9.80,17.82 &  9.30,17.32 &  8.80,16.82 &  8.30,16.32 
&  7.80,15.82 &  7.30,15.32 &  6.80,14.82 &  6.30,14.32 &  5.30,13.32 \\
 15 & 11.32,19.32 & 10.82,18.82 & 10.32,18.32 &  9.82,17.82 &  9.32,17.32 
&  8.82,16.82 &  8.32,16.32 &  7.82,15.82 &  7.32,15.32 &  6.32,14.32 \\
 16 & 12.33,20.80 & 11.83,20.30 & 11.33,19.80 & 10.83,19.30 & 10.33,18.80 
&  9.83,18.30 &  9.33,17.80 &  8.83,17.30 &  8.33,16.80 &  7.33,15.80 \\
 17 & 12.79,21.81 & 12.29,21.31 & 11.79,20.81 & 11.29,20.31 & 10.79,19.81 
& 10.29,19.31 &  9.79,18.81 &  9.29,18.31 &  8.79,17.81 &  7.79,16.81 \\
 18 & 13.81,22.82 & 13.31,22.32 & 12.81,21.82 & 12.31,21.32 & 11.81,20.82 
& 11.31,20.32 & 10.81,19.82 & 10.31,19.32 &  9.81,18.82 &  8.81,17.82 \\
 19 & 14.82,23.82 & 14.32,23.32 & 13.82,22.82 & 13.32,22.32 & 12.82,21.82 
& 12.32,21.32 & 11.82,20.82 & 11.32,20.32 & 10.82,19.82 &  9.82,18.82 \\
 20 & 15.83,25.30 & 15.33,24.80 & 14.83,24.30 & 14.33,23.80 & 13.83,23.30 
& 13.33,22.80 & 12.83,22.30 & 12.33,21.80 & 11.83,21.30 & 10.83,20.30 \\
\end{tabular}
\end{table}

\begin{table}
\caption{68.27\% C.L. intervals for the Poisson signal mean $\mup$, 
for total events observed $n_0$, for known mean background $\mub$
ranging from 6 to 15.}
\label{tab-p68b}
\bigskip
\begin{tabular}{r|cccccccccc} \hline
~$n_0\backslash\mub$~  &    6.0 &    7.0 &    8.0 &    9.0 &   10.0 
&   11.0 &   12.0 &   13.0 &   14.0 &   15.0\\ \hline
  0 & {\it 0.00, 0.18} & {\it 0.00, 0.17} & {\it 0.00, 0.17} 
& {\it 0.00, 0.17} & {\it 0.00, 0.16} & {\it 0.00, 0.16} 
& {\it 0.00, 0.16} & {\it 0.00, 0.16} & {\it 0.00, 0.16} 
& {\it 0.00, 0.15} \\
  1 & 0.00, 0.24 & {\it 0.00, 0.21} & {\it 0.00, 0.20} 
& {\it 0.00, 0.19} & {\it 0.00, 0.18} & {\it 0.00, 0.17} 
& {\it 0.00, 0.17} & {\it 0.00, 0.17} & {\it 0.00, 0.17} 
& {\it 0.00, 0.16} \\
  2 & 0.00, 0.31 & 0.00, 0.27 & 0.00, 0.23 & {\it 0.00, 0.21} 
& {\it 0.00, 0.20} & {\it 0.00, 0.19} & {\it 0.00, 0.19} 
& {\it 0.00, 0.18} & {\it 0.00, 0.18} & {\it 0.00, 0.18} \\
  3 & 0.00, 0.69 & 0.00, 0.42 & 0.00, 0.31 & 0.00, 0.26 & 0.00, 0.23 
& {\it 0.00, 0.22} & {\it 0.00, 0.21} & {\it 0.00, 0.20} 
& {\it 0.00, 0.20} & {\it 0.00, 0.19} \\
  4 & 0.00, 1.22 & 0.00, 0.69 & 0.00, 0.60 & 0.00, 0.38 & 0.00, 0.30 
& 0.00, 0.26 & {\it 0.00, 0.24} & {\it 0.00, 0.23} & {\it 0.00, 0.22} 
& {\it 0.00, 0.21} \\
  5 & 0.00, 1.92 & 0.00, 1.23 & 0.00, 0.99 & 0.00, 0.60 & 0.00, 0.48 
& 0.00, 0.35 & 0.00, 0.29 & 0.00, 0.26 & {\it 0.00, 0.24} 
& {\it 0.00, 0.23} \\
  6 & 0.00, 3.28 & 0.00, 2.38 & 0.00, 1.65 & 0.00, 1.06 & 0.00, 0.63 
& 0.00, 0.53 & 0.00, 0.42 & 0.00, 0.33 & 0.00, 0.29 & {\it 0.00, 0.26} \\
  7 & 0.00, 4.30 & 0.00, 3.30 & 0.00, 2.40 & 0.00, 1.66 & 0.00, 1.07 
& 0.00, 0.88 & 0.00, 0.53 & 0.00, 0.47 & 0.00, 0.38 & 0.00, 0.32 \\
  8 & 0.31, 5.32 & 0.00, 4.32 & 0.00, 3.32 & 0.00, 2.41 & 0.00, 1.67 
& 0.00, 1.46 & 0.00, 0.94 & 0.00, 0.62 & 0.00, 0.48 & 0.00, 0.43 \\
  9 & 0.69, 6.79 & 0.27, 5.79 & 0.00, 4.79 & 0.00, 3.79 & 0.00, 2.87 
& 0.00, 2.10 & 0.00, 1.46 & 0.00, 0.94 & 0.00, 0.78 & 0.00, 0.50 \\
 10 & 1.22, 7.81 & 0.69, 6.81 & 0.23, 5.81 & 0.00, 4.81 & 0.00, 3.81 
& 0.00, 2.89 & 0.00, 2.11 & 0.00, 1.47 & 0.00, 1.03 & 0.00, 0.84 \\
 11 & 1.92, 8.82 & 1.23, 7.82 & 0.60, 6.82 & 0.19, 5.82 & 0.00, 4.82 
& 0.00, 3.82 & 0.00, 2.90 & 0.00, 2.12 & 0.00, 1.54 & 0.00, 1.31 \\
 12 & 2.83,10.29 & 1.94, 9.29 & 1.12, 8.29 & 0.60, 7.29 & 0.12, 6.29 
& 0.00, 5.29 & 0.00, 4.29 & 0.00, 3.36 & 0.00, 2.57 & 0.00, 1.89 \\
 13 & 3.28,11.30 & 2.38,10.30 & 1.65, 9.30 & 1.06, 8.30 & 0.60, 7.30 
& 0.05, 6.30 & 0.00, 5.30 & 0.00, 4.30 & 0.00, 3.37 & 0.00, 2.57 \\
 14 & 4.30,12.32 & 3.30,11.32 & 2.40,10.32 & 1.66, 9.32 & 1.07, 8.32 
& 0.53, 7.32 & 0.00, 6.32 & 0.00, 5.32 & 0.00, 4.32 & 0.00, 3.38 \\
 15 & 5.32,13.32 & 4.32,12.32 & 3.32,11.32 & 2.41,10.32 & 1.67, 9.32 
& 1.00, 8.32 & 0.53, 7.32 & 0.00, 6.32 & 0.00, 5.32 & 0.00, 4.32 \\
 16 & 6.33,14.80 & 5.33,13.80 & 4.33,12.80 & 3.33,11.80 & 2.43,10.80 
& 1.46, 9.80 & 0.94, 8.80 & 0.47, 7.80 & 0.00, 6.80 & 0.00, 5.80 \\
 17 & 6.79,15.81 & 5.79,14.81 & 4.79,13.81 & 3.79,12.81 & 2.87,11.81 
& 2.10,10.81 & 1.46, 9.81 & 0.94, 8.81 & 0.48, 7.81 & 0.00, 6.81 \\
 18 & 7.81,16.82 & 6.81,15.82 & 5.81,14.82 & 4.81,13.82 & 3.81,12.82 
& 2.89,11.82 & 2.11,10.82 & 1.47, 9.82 & 0.93, 8.82 & 0.43, 7.82 \\
 19 & 8.82,17.82 & 7.82,16.82 & 6.82,15.82 & 5.82,14.82 & 4.82,13.82 
& 3.82,12.82 & 2.90,11.82 & 2.12,10.82 & 1.48, 9.82 & 0.84, 8.82 \\
 20 & 9.83,19.30 & 8.83,18.30 & 7.83,17.30 & 6.83,16.30 & 5.83,15.30 
& 4.83,14.30 & 3.83,13.30 & 2.91,12.30 & 2.12,11.30 & 1.31,10.30 \\
\end{tabular}
\end{table}

\begin{table}
\caption{90\% C.L. intervals for the Poisson signal mean $\mup$, 
for total events observed $n_0$, for known mean background $\mub$
ranging from 0 to 5.}
\label{tab-p90a}
\bigskip
\begin{tabular}{r|cccccccccc} \hline
~$n_0\backslash\mub$~  &    0.0 &    0.5 &    1.0 &    1.5 &    2.0 
&    2.5 &    3.0 &    3.5 &    4.0 &    5.0\\ \hline
  0 &  0.00, 2.44 &  0.00, 1.94 &  0.00, 1.61 &  0.00, 1.33 
&  0.00, 1.26 &  0.00, 1.18 &  0.00, 1.08 &  0.00, 1.06 
&  0.00, 1.01 & {\it 0.00, 0.98} \\
  1 &  0.11, 4.36 &  0.00, 3.86 &  0.00, 3.36 &  0.00, 2.91 &  0.00, 2.53 
&  0.00, 2.19 &  0.00, 1.88 &  0.00, 1.59 &  0.00, 1.39 &  0.00, 1.22 \\
  2 &  0.53, 5.91 &  0.03, 5.41 &  0.00, 4.91 &  0.00, 4.41 &  0.00, 3.91 
&  0.00, 3.45 &  0.00, 3.04 &  0.00, 2.67 &  0.00, 2.33 &  0.00, 1.73 \\
  3 &  1.10, 7.42 &  0.60, 6.92 &  0.10, 6.42 &  0.00, 5.92 &  0.00, 5.42 
&  0.00, 4.92 &  0.00, 4.42 &  0.00, 3.95 &  0.00, 3.53 &  0.00, 2.78 \\
  4 &  1.47, 8.60 &  1.17, 8.10 &  0.74, 7.60 &  0.24, 7.10 &  0.00, 6.60 
&  0.00, 6.10 &  0.00, 5.60 &  0.00, 5.10 &  0.00, 4.60 &  0.00, 3.60 \\
  5 &  1.84, 9.99 &  1.53, 9.49 &  1.25, 8.99 &  0.93, 8.49 &  0.43, 7.99 
&  0.00, 7.49 &  0.00, 6.99 &  0.00, 6.49 &  0.00, 5.99 &  0.00, 4.99 \\
  6 &  2.21,11.47 &  1.90,10.97 &  1.61,10.47 &  1.33, 9.97 &  1.08, 9.47 
&  0.65, 8.97 &  0.15, 8.47 &  0.00, 7.97 &  0.00, 7.47 &  0.00, 6.47 \\
  7 &  3.56,12.53 &  3.06,12.03 &  2.56,11.53 &  2.09,11.03 &  1.59,10.53 
&  1.18,10.03 &  0.89, 9.53 &  0.39, 9.03 &  0.00, 8.53 &  0.00, 7.53 \\
  8 &  3.96,13.99 &  3.46,13.49 &  2.96,12.99 &  2.51,12.49 &  2.14,11.99 
&  1.81,11.49 &  1.51,10.99 &  1.06,10.49 &  0.66, 9.99 &  0.00, 8.99 \\
  9 &  4.36,15.30 &  3.86,14.80 &  3.36,14.30 &  2.91,13.80 &  2.53,13.30 
&  2.19,12.80 &  1.88,12.30 &  1.59,11.80 &  1.33,11.30 &  0.43,10.30 \\
 10 &  5.50,16.50 &  5.00,16.00 &  4.50,15.50 &  4.00,15.00 &  3.50,14.50 
&  3.04,14.00 &  2.63,13.50 &  2.27,13.00 &  1.94,12.50 &  1.19,11.50 \\
 11 &  5.91,17.81 &  5.41,17.31 &  4.91,16.81 &  4.41,16.31 &  3.91,15.81 
&  3.45,15.31 &  3.04,14.81 &  2.67,14.31 &  2.33,13.81 &  1.73,12.81 \\
 12 &  7.01,19.00 &  6.51,18.50 &  6.01,18.00 &  5.51,17.50 &  5.01,17.00 
&  4.51,16.50 &  4.01,16.00 &  3.54,15.50 &  3.12,15.00 &  2.38,14.00 \\
 13 &  7.42,20.05 &  6.92,19.55 &  6.42,19.05 &  5.92,18.55 &  5.42,18.05 
&  4.92,17.55 &  4.42,17.05 &  3.95,16.55 &  3.53,16.05 &  2.78,15.05 \\
 14 &  8.50,21.50 &  8.00,21.00 &  7.50,20.50 &  7.00,20.00 &  6.50,19.50 
&  6.00,19.00 &  5.50,18.50 &  5.00,18.00 &  4.50,17.50 &  3.59,16.50 \\
 15 &  9.48,22.52 &  8.98,22.02 &  8.48,21.52 &  7.98,21.02 &  7.48,20.52 
&  6.98,20.02 &  6.48,19.52 &  5.98,19.02 &  5.48,18.52 &  4.48,17.52 \\
 16 &  9.99,23.99 &  9.49,23.49 &  8.99,22.99 &  8.49,22.49 &  7.99,21.99 
&  7.49,21.49 &  6.99,20.99 &  6.49,20.49 &  5.99,19.99 &  4.99,18.99 \\
 17 & 11.04,25.02 & 10.54,24.52 & 10.04,24.02 &  9.54,23.52 &  9.04,23.02 
&  8.54,22.52 &  8.04,22.02 &  7.54,21.52 &  7.04,21.02 &  6.04,20.02 \\
 18 & 11.47,26.16 & 10.97,25.66 & 10.47,25.16 &  9.97,24.66 &  9.47,24.16 
&  8.97,23.66 &  8.47,23.16 &  7.97,22.66 &  7.47,22.16 &  6.47,21.16 \\
 19 & 12.51,27.51 & 12.01,27.01 & 11.51,26.51 & 11.01,26.01 & 10.51,25.51 
& 10.01,25.01 &  9.51,24.51 &  9.01,24.01 &  8.51,23.51 &  7.51,22.51 \\
 20 & 13.55,28.52 & 13.05,28.02 & 12.55,27.52 & 12.05,27.02 & 11.55,26.52 
& 11.05,26.02 & 10.55,25.52 & 10.05,25.02 &  9.55,24.52 &  8.55,23.52 \\
\end{tabular}
\end{table}

\begin{table}
\caption{90\% C.L. intervals for the Poisson signal mean $\mup$, 
for total events observed $n_0$, for known mean background $\mub$
ranging from 6 to 15.}
\label{tab-p90b}
\bigskip
\begin{tabular}{r|cccccccccc} \hline
~$n_0\backslash\mub$~  &    6.0 &    7.0 &    8.0 &    9.0 &   10.0 
&   11.0 &   12.0 &   13.0 &   14.0 &   15.0\\ \hline
  0 & {\it 0.00, 0.97} & {\it 0.00, 0.95} & {\it 0.00, 0.94} 
& {\it 0.00, 0.94} & {\it 0.00, 0.93} & {\it 0.00, 0.93}
 & {\it 0.00, 0.92} & {\it 0.00, 0.92} & {\it 0.00, 0.92} 
& {\it 0.00, 0.92} \\
  1 &  0.00, 1.14 & {\it 0.00, 1.10} & {\it 0.00, 1.07} 
& {\it 0.00, 1.05} & {\it 0.00, 1.03} & {\it 0.00, 1.01} 
& {\it 0.00, 1.00} & {\it 0.00, 0.99} & {\it 0.00, 0.99} 
& {\it 0.00, 0.98} \\
  2 &  0.00, 1.57 &  0.00, 1.38 &  0.00, 1.27 & {\it 0.00, 1.21} 
& {\it 0.00, 1.15} & {\it 0.00, 1.11} & {\it 0.00, 1.09}
 & {\it 0.00, 1.08} & {\it 0.00, 1.06} & {\it 0.00, 1.05} \\
  3 &  0.00, 2.14 &  0.00, 1.75 &  0.00, 1.49 &  0.00, 1.37 
&  0.00, 1.29 & {\it 0.00, 1.24} & {\it 0.00, 1.21} 
& {\it 0.00, 1.18} & {\it 0.00, 1.15} & {\it 0.00, 1.14} \\
  4 &  0.00, 2.83 &  0.00, 2.56 &  0.00, 1.98 &  0.00, 1.82 
&  0.00, 1.57 &  0.00, 1.45 & {\it 0.00, 1.37} & {\it 0.00, 1.31}
 & {\it 0.00, 1.27} & {\it 0.00, 1.24} \\
  5 &  0.00, 4.07 &  0.00, 3.28 &  0.00, 2.60 &  0.00, 2.38 
&  0.00, 1.85 &  0.00, 1.70 &  0.00, 1.58 &  0.00, 1.48 
& {\it 0.00, 1.39} & {\it 0.00, 1.32} \\
  6 &  0.00, 5.47 &  0.00, 4.54 &  0.00, 3.73 &  0.00, 3.02 
&  0.00, 2.40 &  0.00, 2.21 &  0.00, 1.86 &  0.00, 1.67 
&  0.00, 1.55 & {\it 0.00, 1.47} \\
  7 &  0.00, 6.53 &  0.00, 5.53 &  0.00, 4.58 &  0.00, 3.77 &  0.00, 3.26 
&  0.00, 2.81 &  0.00, 2.23 &  0.00, 2.07 &  0.00, 1.86 &  0.00, 1.69 \\
  8 &  0.00, 7.99 &  0.00, 6.99 &  0.00, 5.99 &  0.00, 5.05 &  0.00, 4.22 
&  0.00, 3.49 &  0.00, 2.83 &  0.00, 2.62 &  0.00, 2.11 &  0.00, 1.95 \\
  9 &  0.00, 9.30 &  0.00, 8.30 &  0.00, 7.30 &  0.00, 6.30 &  0.00, 5.30 
&  0.00, 4.30 &  0.00, 3.93 &  0.00, 3.25 &  0.00, 2.64 &  0.00, 2.45 \\
 10 &  0.22,10.50 &  0.00, 9.50 &  0.00, 8.50 &  0.00, 7.50 &  0.00, 6.50 
&  0.00, 5.56 &  0.00, 4.71 &  0.00, 3.95 &  0.00, 3.27 &  0.00, 3.00 \\
 11 &  1.01,11.81 &  0.02,10.81 &  0.00, 9.81 &  0.00, 8.81 &  0.00, 7.81 
&  0.00, 6.81 &  0.00, 5.81 &  0.00, 4.81 &  0.00, 4.39 &  0.00, 3.69 \\
 12 &  1.57,13.00 &  0.83,12.00 &  0.00,11.00 &  0.00,10.00 &  0.00, 9.00 
&  0.00, 8.00 &  0.00, 7.00 &  0.00, 6.05 &  0.00, 5.19 &  0.00, 4.42 \\
 13 &  2.14,14.05 &  1.50,13.05 &  0.65,12.05 &  0.00,11.05 &  0.00,10.05 
&  0.00, 9.05 &  0.00, 8.05 &  0.00, 7.05 &  0.00, 6.08 &  0.00, 5.22 \\
 14 &  2.83,15.50 &  2.13,14.50 &  1.39,13.50 &  0.47,12.50 &  0.00,11.50 
&  0.00,10.50 &  0.00, 9.50 &  0.00, 8.50 &  0.00, 7.50 &  0.00, 6.55 \\
 15 &  3.48,16.52 &  2.56,15.52 &  1.98,14.52 &  1.26,13.52 &  0.30,12.52 
&  0.00,11.52 &  0.00,10.52 &  0.00, 9.52 &  0.00, 8.52 &  0.00, 7.52 \\
 16 &  4.07,17.99 &  3.28,16.99 &  2.60,15.99 &  1.82,14.99 &  1.13,13.99 
&  0.14,12.99 &  0.00,11.99 &  0.00,10.99 &  0.00, 9.99 &  0.00, 8.99 \\
 17 &  5.04,19.02 &  4.11,18.02 &  3.32,17.02 &  2.38,16.02 &  1.81,15.02 
&  0.98,14.02 &  0.00,13.02 &  0.00,12.02 &  0.00,11.02 &  0.00,10.02 \\
 18 &  5.47,20.16 &  4.54,19.16 &  3.73,18.16 &  3.02,17.16 &  2.40,16.16 
&  1.70,15.16 &  0.82,14.16 &  0.00,13.16 &  0.00,12.16 &  0.00,11.16 \\
 19 &  6.51,21.51 &  5.51,20.51 &  4.58,19.51 &  3.77,18.51 &  3.05,17.51 
&  2.21,16.51 &  1.58,15.51 &  0.67,14.51 &  0.00,13.51 &  0.00,12.51 \\
 20 &  7.55,22.52 &  6.55,21.52 &  5.55,20.52 &  4.55,19.52 &  3.55,18.52 
&  2.81,17.52 &  2.23,16.52 &  1.48,15.52 &  0.53,14.52 &  0.00,13.52 \\
\end{tabular}
\end{table}

\begin{table}
\caption{95\% C.L. intervals for the Poisson signal mean $\mup$, 
for total events observed $n_0$, for known mean background $\mub$
ranging from 0 to 5.}
\label{tab-p95a}
\bigskip
\begin{tabular}{r|cccccccccc} \hline
~$n_0\backslash\mub$~  &    0.0 &    0.5 &    1.0 &    1.5 &    2.0 
&    2.5 &    3.0 &    3.5 &    4.0 &    5.0\\ \hline
  0 &  0.00, 3.09 &  0.00, 2.63 &  0.00, 2.33 &  0.00, 2.05 &  0.00, 1.78 
&  0.00, 1.78 &  0.00, 1.63 &  0.00, 1.63 &  0.00, 1.57 
& {\it 0.00, 1.54} \\
  1 &  0.05, 5.14 &  0.00, 4.64 &  0.00, 4.14 &  0.00, 3.69 &  0.00, 3.30
 &  0.00, 2.95 &  0.00, 2.63 &  0.00, 2.33 &  0.00, 2.08 &  0.00, 1.88 \\
  2 &  0.36, 6.72 &  0.00, 6.22 &  0.00, 5.72 &  0.00, 5.22 &  0.00, 4.72 
&  0.00, 4.25 &  0.00, 3.84 &  0.00, 3.46 &  0.00, 3.11 &  0.00, 2.49 \\
  3 &  0.82, 8.25 &  0.32, 7.75 &  0.00, 7.25 &  0.00, 6.75 &  0.00, 6.25 
&  0.00, 5.75 &  0.00, 5.25 &  0.00, 4.78 &  0.00, 4.35 &  0.00, 3.58 \\
  4 &  1.37, 9.76 &  0.87, 9.26 &  0.37, 8.76 &  0.00, 8.26 &  0.00, 7.76 
&  0.00, 7.26 &  0.00, 6.76 &  0.00, 6.26 &  0.00, 5.76 &  0.00, 4.84 \\
  5 &  1.84,11.26 &  1.47,10.76 &  0.97,10.26 &  0.47, 9.76 &  0.00, 9.26 
&  0.00, 8.76 &  0.00, 8.26 &  0.00, 7.76 &  0.00, 7.26 &  0.00, 6.26 \\
  6 &  2.21,12.75 &  1.90,12.25 &  1.61,11.75 &  1.11,11.25 &  0.61,10.75 
&  0.11,10.25 &  0.00, 9.75 &  0.00, 9.25 &  0.00, 8.75 &  0.00, 7.75 \\
  7 &  2.58,13.81 &  2.27,13.31 &  1.97,12.81 &  1.69,12.31 &  1.29,11.81 
&  0.79,11.31 &  0.29,10.81 &  0.00,10.31 &  0.00, 9.81 &  0.00, 8.81 \\
  8 &  2.94,15.29 &  2.63,14.79 &  2.33,14.29 &  2.05,13.79 &  1.78,13.29 
&  1.48,12.79 &  0.98,12.29 &  0.48,11.79 &  0.00,11.29 &  0.00,10.29 \\
  9 &  4.36,16.77 &  3.86,16.27 &  3.36,15.77 &  2.91,15.27 &  2.46,14.77 
&  1.96,14.27 &  1.62,13.77 &  1.20,13.27 &  0.70,12.77 &  0.00,11.77 \\
 10 &  4.75,17.82 &  4.25,17.32 &  3.75,16.82 &  3.30,16.32 &  2.92,15.82 
&  2.57,15.32 &  2.25,14.82 &  1.82,14.32 &  1.43,13.82 &  0.43,12.82 \\
 11 &  5.14,19.29 &  4.64,18.79 &  4.14,18.29 &  3.69,17.79 &  3.30,17.29 
&  2.95,16.79 &  2.63,16.29 &  2.33,15.79 &  2.04,15.29 &  1.17,14.29 \\
 12 &  6.32,20.34 &  5.82,19.84 &  5.32,19.34 &  4.82,18.84 &  4.32,18.34 
&  3.85,17.84 &  3.44,17.34 &  3.06,16.84 &  2.69,16.34 &  1.88,15.34 \\
 13 &  6.72,21.80 &  6.22,21.30 &  5.72,20.80 &  5.22,20.30 &  4.72,19.80 
&  4.25,19.30 &  3.84,18.80 &  3.46,18.30 &  3.11,17.80 &  2.47,16.80 \\
 14 &  7.84,22.94 &  7.34,22.44 &  6.84,21.94 &  6.34,21.44 &  5.84,20.94 
&  5.34,20.44 &  4.84,19.94 &  4.37,19.44 &  3.94,18.94 &  3.10,17.94 \\
 15 &  8.25,24.31 &  7.75,23.81 &  7.25,23.31 &  6.75,22.81 &  6.25,22.31 
&  5.75,21.81 &  5.25,21.31 &  4.78,20.81 &  4.35,20.31 &  3.58,19.31 \\
 16 &  9.34,25.40 &  8.84,24.90 &  8.34,24.40 &  7.84,23.90 &  7.34,23.40 
&  6.84,22.90 &  6.34,22.40 &  5.84,21.90 &  5.34,21.40 &  4.43,20.40 \\
 17 &  9.76,26.81 &  9.26,26.31 &  8.76,25.81 &  8.26,25.31 &  7.76,24.81 
&  7.26,24.31 &  6.76,23.81 &  6.26,23.31 &  5.76,22.81 &  4.84,21.81 \\
 18 & 10.84,27.84 & 10.34,27.34 &  9.84,26.84 &  9.34,26.34 &  8.84,25.84 
&  8.34,25.34 &  7.84,24.84 &  7.34,24.34 &  6.84,23.84 &  5.84,22.84 \\
 19 & 11.26,29.31 & 10.76,28.81 & 10.26,28.31 &  9.76,27.81 &  9.26,27.31 
&  8.76,26.81 &  8.26,26.31 &  7.76,25.81 &  7.26,25.31 &  6.26,24.31 \\
 20 & 12.33,30.33 & 11.83,29.83 & 11.33,29.33 & 10.83,28.83 & 10.33,28.33
 &  9.83,27.83 &  9.33,27.33 &  8.83,26.83 &  8.33,26.33 &  7.33,25.33 \\
\end{tabular}
\end{table}

\begin{table}
\caption{95\% C.L. intervals for the Poisson signal mean $\mup$, 
for total events observed $n_0$, for known mean background $\mub$
ranging from 6 to 15.}
\label{tab-p95b}
\bigskip
\begin{tabular}{r|cccccccccc} \hline
~$n_0\backslash\mub$~  &    6.0 &    7.0 &    8.0 &    9.0 &   10.0 
&   11.0 &   12.0 &   13.0 &   14.0 &   15.0\\ \hline
  0 & {\it 0.00, 1.52} & {\it 0.00, 1.51} & {\it 0.00, 1.50} 
& {\it 0.00, 1.49} & {\it 0.00, 1.49} & {\it 0.00, 1.48} 
& {\it 0.00, 1.48} & {\it 0.00, 1.48} & {\it 0.00, 1.47} 
& {\it 0.00, 1.47} \\
  1 &  0.00, 1.78 & {\it 0.00, 1.73} & {\it 0.00, 1.69} 
& {\it 0.00, 1.66} & {\it 0.00, 1.64} & {\it 0.00, 1.61}
 & {\it 0.00, 1.60} & {\it 0.00, 1.59} & {\it 0.00, 1.58} 
& {\it 0.00, 1.56} \\
  2 &  0.00, 2.28 &  0.00, 2.11 &  0.00, 1.98 & {\it 0.00, 1.86} 
& {\it 0.00, 1.81} & {\it 0.00, 1.77} & {\it 0.00, 1.74} 
& {\it 0.00, 1.72} & {\it 0.00, 1.70} & {\it 0.00, 1.67} \\
  3 &  0.00, 2.91 &  0.00, 2.69 &  0.00, 2.37 &  0.00, 2.17 
&  0.00, 2.06 & {\it 0.00, 1.98} & {\it 0.00, 1.93} 
& {\it 0.00, 1.89} & {\it 0.00, 1.82} & {\it 0.00, 1.80} \\
  4 &  0.00, 4.05 &  0.00, 3.35 &  0.00, 3.01 &  0.00, 2.54 
&  0.00, 2.37 &  0.00, 2.23 & {\it 0.00, 2.11} & {\it 0.00, 2.04}
 & {\it 0.00, 1.99} & {\it 0.00, 1.95} \\
  5 &  0.00, 5.33 &  0.00, 4.52 &  0.00, 3.79 &  0.00, 3.15 
&  0.00, 2.94 &  0.00, 2.65 &  0.00, 2.43 &  0.00, 2.30 
& {\it 0.00, 2.20} & {\it 0.00, 2.13} \\
  6 &  0.00, 6.75 &  0.00, 5.82 &  0.00, 4.99 &  0.00, 4.24 
&  0.00, 3.57 &  0.00, 3.14 &  0.00, 2.78 &  0.00, 2.62 
&  0.00, 2.48 & {\it 0.00, 2.35} \\
  7 &  0.00, 7.81 &  0.00, 6.81 &  0.00, 5.87 &  0.00, 5.03 &  0.00, 4.28 
&  0.00, 4.00 &  0.00, 3.37 &  0.00, 3.15 &  0.00, 2.79 &  0.00, 2.59 \\
  8 &  0.00, 9.29 &  0.00, 8.29 &  0.00, 7.29 &  0.00, 6.35 &  0.00, 5.50 
&  0.00, 4.73 &  0.00, 4.03 &  0.00, 3.79 &  0.00, 3.20 &  0.00, 3.02 \\
  9 &  0.00,10.77 &  0.00, 9.77 &  0.00, 8.77 &  0.00, 7.77 &  0.00, 6.82 
&  0.00, 5.96 &  0.00, 5.18 &  0.00, 4.47 &  0.00, 3.81 &  0.00, 3.60 \\
 10 &  0.00,11.82 &  0.00,10.82 &  0.00, 9.82 &  0.00, 8.82 &  0.00, 7.82 
&  0.00, 6.87 &  0.00, 6.00 &  0.00, 5.21 &  0.00, 4.59 &  0.00, 4.24 \\
 11 &  0.17,13.29 &  0.00,12.29 &  0.00,11.29 &  0.00,10.29 &  0.00, 9.29
 &  0.00, 8.29 &  0.00, 7.34 &  0.00, 6.47 &  0.00, 5.67 &  0.00, 4.93 \\
 12 &  0.92,14.34 &  0.00,13.34 &  0.00,12.34 &  0.00,11.34 &  0.00,10.34 
&  0.00, 9.34 &  0.00, 8.34 &  0.00, 7.37 &  0.00, 6.50 &  0.00, 5.70 \\
 13 &  1.68,15.80 &  0.69,14.80 &  0.00,13.80 &  0.00,12.80 &  0.00,11.80 
&  0.00,10.80 &  0.00, 9.80 &  0.00, 8.80 &  0.00, 7.85 &  0.00, 6.96 \\
 14 &  2.28,16.94 &  1.46,15.94 &  0.46,14.94 &  0.00,13.94 &  0.00,12.94 
&  0.00,11.94 &  0.00,10.94 &  0.00, 9.94 &  0.00, 8.94 &  0.00, 7.94 \\
 15 &  2.91,18.31 &  2.11,17.31 &  1.25,16.31 &  0.25,15.31 &  0.00,14.31 
&  0.00,13.31 &  0.00,12.31 &  0.00,11.31 &  0.00,10.31 &  0.00, 9.31 \\
 16 &  3.60,19.40 &  2.69,18.40 &  1.98,17.40 &  1.04,16.40 &  0.04,15.40 
&  0.00,14.40 &  0.00,13.40 &  0.00,12.40 &  0.00,11.40 &  0.00,10.40 \\
 17 &  4.05,20.81 &  3.35,19.81 &  2.63,18.81 &  1.83,17.81 &  0.83,16.81 
&  0.00,15.81 &  0.00,14.81 &  0.00,13.81 &  0.00,12.81 &  0.00,11.81 \\
 18 &  4.91,21.84 &  4.11,20.84 &  3.18,19.84 &  2.53,18.84 &  1.63,17.84 
&  0.63,16.84 &  0.00,15.84 &  0.00,14.84 &  0.00,13.84 &  0.00,12.84 \\
 19 &  5.33,23.31 &  4.52,22.31 &  3.79,21.31 &  3.15,20.31 &  2.37,19.31 
&  1.44,18.31 &  0.44,17.31 &  0.00,16.31 &  0.00,15.31 &  0.00,14.31 \\
 20 &  6.33,24.33 &  5.39,23.33 &  4.57,22.33 &  3.82,21.33 &  2.94,20.33 
&  2.23,19.33 &  1.25,18.33 &  0.25,17.33 &  0.00,16.33 &  0.00,15.33 \\
\end{tabular}
\end{table}

\begin{table}
\caption{99\% C.L. intervals for the Poisson signal mean $\mup$, 
for total events observed $n_0$, for known mean background $\mub$
ranging from 0 to 5.}
\label{tab-p99a}
\bigskip
\begin{tabular}{r|cccccccccc} \hline
~$n_0\backslash\mub$~  &    0.0 &    0.5 &    1.0 &    1.5 &    2.0 
&    2.5 &    3.0 &    3.5 &    4.0 &    5.0\\ \hline
  0 &  0.00, 4.74 &  0.00, 4.24 &  0.00, 3.80 &  0.00, 3.50 &  0.00, 3.26
 &  0.00, 3.26 &  0.00, 3.05 &  0.00, 3.05 &  0.00, 2.98 
& {\it 0.00, 2.94} \\
  1 &  0.01, 6.91 &  0.00, 6.41 &  0.00, 5.91 &  0.00, 5.41 &  0.00, 4.91 
&  0.00, 4.48 &  0.00, 4.14 &  0.00, 4.09 &  0.00, 3.89 &  0.00, 3.59 \\
  2 &  0.15, 8.71 &  0.00, 8.21 &  0.00, 7.71 &  0.00, 7.21 &  0.00, 6.71 
&  0.00, 6.24 &  0.00, 5.82 &  0.00, 5.42 &  0.00, 5.06 &  0.00, 4.37 \\
  3 &  0.44,10.47 &  0.00, 9.97 &  0.00, 9.47 &  0.00, 8.97 &  0.00, 8.47 
&  0.00, 7.97 &  0.00, 7.47 &  0.00, 6.97 &  0.00, 6.47 &  0.00, 5.57 \\
  4 &  0.82,12.23 &  0.32,11.73 &  0.00,11.23 &  0.00,10.73 &  0.00,10.23 
&  0.00, 9.73 &  0.00, 9.23 &  0.00, 8.73 &  0.00, 8.23 &  0.00, 7.30 \\
  5 &  1.28,13.75 &  0.78,13.25 &  0.28,12.75 &  0.00,12.25 &  0.00,11.75 
&  0.00,11.25 &  0.00,10.75 &  0.00,10.25 &  0.00, 9.75 &  0.00, 8.75 \\
  6 &  1.79,15.27 &  1.29,14.77 &  0.79,14.27 &  0.29,13.77 &  0.00,13.27 
&  0.00,12.77 &  0.00,12.27 &  0.00,11.77 &  0.00,11.27 &  0.00,10.27 \\
  7 &  2.33,16.77 &  1.83,16.27 &  1.33,15.77 &  0.83,15.27 &  0.33,14.77 
&  0.00,14.27 &  0.00,13.77 &  0.00,13.27 &  0.00,12.77 &  0.00,11.77 \\
  8 &  2.91,18.27 &  2.41,17.77 &  1.91,17.27 &  1.41,16.77 &  0.91,16.27 
&  0.41,15.77 &  0.00,15.27 &  0.00,14.77 &  0.00,14.27 &  0.00,13.27 \\
  9 &  3.31,19.46 &  3.00,18.96 &  2.51,18.46 &  2.01,17.96 &  1.51,17.46 
&  1.01,16.96 &  0.51,16.46 &  0.01,15.96 &  0.00,15.46 &  0.00,14.46 \\
 10 &  3.68,20.83 &  3.37,20.33 &  3.07,19.83 &  2.63,19.33 &  2.13,18.83 
&  1.63,18.33 &  1.13,17.83 &  0.63,17.33 &  0.13,16.83 &  0.00,15.83 \\
 11 &  4.05,22.31 &  3.73,21.81 &  3.43,21.31 &  3.14,20.81 &  2.77,20.31 
&  2.27,19.81 &  1.77,19.31 &  1.27,18.81 &  0.77,18.31 &  0.00,17.31 \\
 12 &  4.41,23.80 &  4.10,23.30 &  3.80,22.80 &  3.50,22.30 &  3.22,21.80 
&  2.93,21.30 &  2.43,20.80 &  1.93,20.30 &  1.43,19.80 &  0.43,18.80 \\
 13 &  5.83,24.92 &  5.33,24.42 &  4.83,23.92 &  4.33,23.42 &  3.83,22.92 
&  3.33,22.42 &  3.02,21.92 &  2.60,21.42 &  2.10,20.92 &  1.10,19.92 \\
 14 &  6.31,26.33 &  5.81,25.83 &  5.31,25.33 &  4.86,24.83 &  4.46,24.33 
&  4.10,23.83 &  3.67,23.33 &  3.17,22.83 &  2.78,22.33 &  1.78,21.33 \\
 15 &  6.70,27.81 &  6.20,27.31 &  5.70,26.81 &  5.24,26.31 &  4.84,25.81 
&  4.48,25.31 &  4.14,24.81 &  3.82,24.31 &  3.42,23.81 &  2.48,22.81 \\
 16 &  7.76,28.85 &  7.26,28.35 &  6.76,27.85 &  6.26,27.35 &  5.76,26.85 
&  5.26,26.35 &  4.76,25.85 &  4.26,25.35 &  3.89,24.85 &  3.15,23.85 \\
 17 &  8.32,30.33 &  7.82,29.83 &  7.32,29.33 &  6.82,28.83 &  6.32,28.33 
&  5.85,27.83 &  5.42,27.33 &  5.03,26.83 &  4.67,26.33 &  3.73,25.33 \\
 18 &  8.71,31.81 &  8.21,31.31 &  7.71,30.81 &  7.21,30.31 &  6.71,29.81 
&  6.24,29.31 &  5.82,28.81 &  5.42,28.31 &  5.06,27.81 &  4.37,26.81 \\
 19 &  9.88,32.85 &  9.38,32.35 &  8.88,31.85 &  8.38,31.35 &  7.88,30.85 
&  7.38,30.35 &  6.88,29.85 &  6.40,29.35 &  5.97,28.85 &  5.01,27.85 \\
 20 & 10.28,34.32 &  9.78,33.82 &  9.28,33.32 &  8.78,32.82 &  8.28,32.32 
&  7.78,31.82 &  7.28,31.32 &  6.81,30.82 &  6.37,30.32 &  5.57,29.32 \\
\end{tabular}
\end{table}

\begin{table}
\caption{99\% C.L. intervals for the Poisson signal mean $\mup$, 
for total events observed $n_0$, for known mean background $\mub$
ranging from 6 to 15.}
\label{tab-p99b}
\bigskip
\begin{tabular}{r|cccccccccc} \hline
~$n_0\backslash\mub$~  &    6.0 &    7.0 &    8.0 &    9.0 &   10.0 
&   11.0 &   12.0 &   13.0 &   14.0 &   15.0\\ \hline
  0 & {\it 0.00, 2.91} & {\it 0.00, 2.90} & {\it 0.00, 2.89} 
& {\it 0.00, 2.88} & {\it 0.00, 2.88} & {\it 0.00, 2.87} 
& {\it 0.00, 2.87} & {\it 0.00, 2.86} & {\it 0.00, 2.86} 
& {\it 0.00, 2.86} \\
  1 &  0.00, 3.42 & {\it 0.00, 3.31} & {\it 0.00, 3.21} 
& {\it 0.00, 3.18} & {\it 0.00, 3.15} & {\it 0.00, 3.11} 
& {\it 0.00, 3.09} & {\it 0.00, 3.07} & {\it 0.00, 3.06} 
& {\it 0.00, 3.03} \\
  2 &  0.00, 4.13 &  0.00, 3.89 &  0.00, 3.70 & {\it 0.00, 3.56} 
& {\it 0.00, 3.44} & {\it 0.00, 3.39} & {\it 0.00, 3.35} 
& {\it 0.00, 3.32} & {\it 0.00, 3.26} & {\it 0.00, 3.23} \\
  3 &  0.00, 5.25 &  0.00, 4.59 &  0.00, 4.35 &  0.00, 4.06 
&  0.00, 3.89 & {\it 0.00, 3.77} & {\it 0.00, 3.65}
& {\it 0.00, 3.56} & {\it 0.00, 3.51} & {\it 0.00, 3.47} \\
  4 &  0.00, 6.47 &  0.00, 5.73 &  0.00, 5.04 &  0.00, 4.79 
&  0.00, 4.39 &  0.00, 4.17 & {\it 0.00, 4.02} & {\it 0.00, 3.91} 
& {\it 0.00, 3.82} & {\it 0.00, 3.74} \\
  5 &  0.00, 7.81 &  0.00, 6.97 &  0.00, 6.21 &  0.00, 5.50 
&  0.00, 5.17 &  0.00, 4.67 &  0.00, 4.42 &  0.00, 4.24 
& {\it 0.00, 4.11} & {\it 0.00, 4.01} \\
  6 &  0.00, 9.27 &  0.00, 8.32 &  0.00, 7.47 &  0.00, 6.68 
&  0.00, 5.96 &  0.00, 5.46 &  0.00, 5.05 &  0.00, 4.83 
&  0.00, 4.63 & {\it 0.00, 4.44} \\
  7 &  0.00,10.77 &  0.00, 9.77 &  0.00, 8.82 &  0.00, 7.95 &  0.00, 7.16 
&  0.00, 6.42 &  0.00, 5.73 &  0.00, 5.48 &  0.00, 5.12 &  0.00, 4.82 \\
  8 &  0.00,12.27 &  0.00,11.27 &  0.00,10.27 &  0.00, 9.31 &  0.00, 8.44 
&  0.00, 7.63 &  0.00, 6.88 &  0.00, 6.18 &  0.00, 5.83 &  0.00, 5.29 \\
  9 &  0.00,13.46 &  0.00,12.46 &  0.00,11.46 &  0.00,10.46 &  0.00, 9.46 
&  0.00, 8.50 &  0.00, 7.69 &  0.00, 7.34 &  0.00, 6.62 &  0.00, 5.95 \\
 10 &  0.00,14.83 &  0.00,13.83 &  0.00,12.83 &  0.00,11.83 &  0.00,10.83 
&  0.00, 9.87 &  0.00, 8.98 &  0.00, 8.16 &  0.00, 7.39 &  0.00, 7.07 \\
 11 &  0.00,16.31 &  0.00,15.31 &  0.00,14.31 &  0.00,13.31 &  0.00,12.31 
&  0.00,11.31 &  0.00,10.35 &  0.00, 9.46 &  0.00, 8.63 &  0.00, 7.84 \\
 12 &  0.00,17.80 &  0.00,16.80 &  0.00,15.80 &  0.00,14.80 &  0.00,13.80 
&  0.00,12.80 &  0.00,11.80 &  0.00,10.83 &  0.00, 9.94 &  0.00, 9.09 \\
 13 &  0.10,18.92 &  0.00,17.92 &  0.00,16.92 &  0.00,15.92 &  0.00,14.92 
&  0.00,13.92 &  0.00,12.92 &  0.00,11.92 &  0.00,10.92 &  0.00, 9.98 \\
 14 &  0.78,20.33 &  0.00,19.33 &  0.00,18.33 &  0.00,17.33 &  0.00,16.33 
&  0.00,15.33 &  0.00,14.33 &  0.00,13.33 &  0.00,12.33 &  0.00,11.36 \\
 15 &  1.48,21.81 &  0.48,20.81 &  0.00,19.81 &  0.00,18.81 &  0.00,17.81 
&  0.00,16.81 &  0.00,15.81 &  0.00,14.81 &  0.00,13.81 &  0.00,12.81 \\
 16 &  2.18,22.85 &  1.18,21.85 &  0.18,20.85 &  0.00,19.85 &  0.00,18.85 
&  0.00,17.85 &  0.00,16.85 &  0.00,15.85 &  0.00,14.85 &  0.00,13.85 \\
 17 &  2.89,24.33 &  1.89,23.33 &  0.89,22.33 &  0.00,21.33 &  0.00,20.33 
&  0.00,19.33 &  0.00,18.33 &  0.00,17.33 &  0.00,16.33 &  0.00,15.33 \\
 18 &  3.53,25.81 &  2.62,24.81 &  1.62,23.81 &  0.62,22.81 &  0.00,21.81 
&  0.00,20.81 &  0.00,19.81 &  0.00,18.81 &  0.00,17.81 &  0.00,16.81 \\
 19 &  4.13,26.85 &  3.31,25.85 &  2.35,24.85 &  1.35,23.85 &  0.35,22.85 
&  0.00,21.85 &  0.00,20.85 &  0.00,19.85 &  0.00,18.85 &  0.00,17.85 \\
 20 &  4.86,28.32 &  3.93,27.32 &  3.08,26.32 &  2.08,25.32 &  1.08,24.32 
&  0.08,23.32 &  0.00,22.32 &  0.00,21.32 &  0.00,20.32 &  0.00,19.32 \\
\end{tabular}
\end{table}
} 

{\squeezetable
\narrowtext
\begin{table}
\caption{Our confidence intervals for the mean $\mup$ of a Gaussian,
constrained to be non-negative, as a function of the measured mean
$x_0$, for commonly used confidence levels.  Italicized intervals
corresponds to cases where the goodness-of-fit probability
(Sec.~\ref{goodness-of-fit}) is less than 1\%. 
All numbers are in units of $\sigma$. }
\label{tab-gauss-new}
\bigskip
\begin{tabular}{dcccc} \hline
 $x_0$ & 68.27\% C.L.  & 90\% C.L.  & 95\% C.L.  & 99\% C.L.  \\ \hline
  -3.0 & {\it  0.00, 0.04 } & {\it  0.00, 0.26 } & {\it  0.00, 0.42 } 
& {\it  0.00, 0.80 } \\
  -2.9 & {\it  0.00, 0.04 } & {\it  0.00, 0.27 } & {\it  0.00, 0.44 } 
& {\it  0.00, 0.82 } \\
  -2.8 & {\it  0.00, 0.04 } & {\it  0.00, 0.28 } & {\it  0.00, 0.45 } 
& {\it  0.00, 0.84 } \\
  -2.7 & {\it  0.00, 0.04 } & {\it  0.00, 0.29 } & {\it  0.00, 0.47 } 
& {\it  0.00, 0.87 } \\
  -2.6 & {\it  0.00, 0.05 } & {\it  0.00, 0.30 } & {\it  0.00, 0.48 } 
& {\it  0.00, 0.89 } \\
  -2.5 & {\it  0.00, 0.05 } & {\it  0.00, 0.32 } & {\it  0.00, 0.50 } 
& {\it  0.00, 0.92 } \\
  -2.4 & {\it  0.00, 0.05 } & {\it  0.00, 0.33 } & {\it  0.00, 0.52 } 
& {\it  0.00, 0.95 } \\
  -2.3 &  0.00, 0.05 &  0.00, 0.34 &  0.00, 0.54 &  0.00, 0.99 \\
  -2.2 &  0.00, 0.06 &  0.00, 0.36 &  0.00, 0.56 &  0.00, 1.02 \\
  -2.1 &  0.00, 0.06 &  0.00, 0.38 &  0.00, 0.59 &  0.00, 1.06 \\
  -2.0 &  0.00, 0.07 &  0.00, 0.40 &  0.00, 0.62 &  0.00, 1.10 \\
  -1.9 &  0.00, 0.08 &  0.00, 0.43 &  0.00, 0.65 &  0.00, 1.14 \\
  -1.8 &  0.00, 0.09 &  0.00, 0.45 &  0.00, 0.68 &  0.00, 1.19 \\
  -1.7 &  0.00, 0.10 &  0.00, 0.48 &  0.00, 0.72 &  0.00, 1.24 \\
  -1.6 &  0.00, 0.11 &  0.00, 0.52 &  0.00, 0.76 &  0.00, 1.29 \\
  -1.5 &  0.00, 0.13 &  0.00, 0.56 &  0.00, 0.81 &  0.00, 1.35 \\
  -1.4 &  0.00, 0.15 &  0.00, 0.60 &  0.00, 0.86 &  0.00, 1.41 \\
  -1.3 &  0.00, 0.17 &  0.00, 0.64 &  0.00, 0.91 &  0.00, 1.47 \\
  -1.2 &  0.00, 0.20 &  0.00, 0.70 &  0.00, 0.97 &  0.00, 1.54 \\
  -1.1 &  0.00, 0.23 &  0.00, 0.75 &  0.00, 1.04 &  0.00, 1.61 \\
  -1.0 &  0.00, 0.27 &  0.00, 0.81 &  0.00, 1.10 &  0.00, 1.68 \\
  -0.9 &  0.00, 0.32 &  0.00, 0.88 &  0.00, 1.17 &  0.00, 1.76 \\
  -0.8 &  0.00, 0.37 &  0.00, 0.95 &  0.00, 1.25 &  0.00, 1.84 \\
  -0.7 &  0.00, 0.43 &  0.00, 1.02 &  0.00, 1.33 &  0.00, 1.93 \\
  -0.6 &  0.00, 0.49 &  0.00, 1.10 &  0.00, 1.41 &  0.00, 2.01 \\
  -0.5 &  0.00, 0.56 &  0.00, 1.18 &  0.00, 1.49 &  0.00, 2.10 \\
  -0.4 &  0.00, 0.64 &  0.00, 1.27 &  0.00, 1.58 &  0.00, 2.19 \\
  -0.3 &  0.00, 0.72 &  0.00, 1.36 &  0.00, 1.67 &  0.00, 2.28 \\
  -0.2 &  0.00, 0.81 &  0.00, 1.45 &  0.00, 1.77 &  0.00, 2.38 \\
  -0.1 &  0.00, 0.90 &  0.00, 1.55 &  0.00, 1.86 &  0.00, 2.48 \\
   0.0 &  0.00, 1.00 &  0.00, 1.64 &  0.00, 1.96 &  0.00, 2.58 \\
   0.1 &  0.00, 1.10 &  0.00, 1.74 &  0.00, 2.06 &  0.00, 2.68 \\
   0.2 &  0.00, 1.20 &  0.00, 1.84 &  0.00, 2.16 &  0.00, 2.78 \\
   0.3 &  0.00, 1.30 &  0.00, 1.94 &  0.00, 2.26 &  0.00, 2.88 \\
   0.4 &  0.00, 1.40 &  0.00, 2.04 &  0.00, 2.36 &  0.00, 2.98 \\
   0.5 &  0.02, 1.50 &  0.00, 2.14 &  0.00, 2.46 &  0.00, 3.08 \\
   0.6 &  0.07, 1.60 &  0.00, 2.24 &  0.00, 2.56 &  0.00, 3.18 \\
   0.7 &  0.11, 1.70 &  0.00, 2.34 &  0.00, 2.66 &  0.00, 3.28 \\
   0.8 &  0.15, 1.80 &  0.00, 2.44 &  0.00, 2.76 &  0.00, 3.38 \\
   0.9 &  0.19, 1.90 &  0.00, 2.54 &  0.00, 2.86 &  0.00, 3.48 \\
   1.0 &  0.24, 2.00 &  0.00, 2.64 &  0.00, 2.96 &  0.00, 3.58 \\
   1.1 &  0.30, 2.10 &  0.00, 2.74 &  0.00, 3.06 &  0.00, 3.68 \\
   1.2 &  0.35, 2.20 &  0.00, 2.84 &  0.00, 3.16 &  0.00, 3.78 \\
   1.3 &  0.42, 2.30 &  0.02, 2.94 &  0.00, 3.26 &  0.00, 3.88 \\
   1.4 &  0.49, 2.40 &  0.12, 3.04 &  0.00, 3.36 &  0.00, 3.98 \\
   1.5 &  0.56, 2.50 &  0.22, 3.14 &  0.00, 3.46 &  0.00, 4.08 \\
   1.6 &  0.64, 2.60 &  0.31, 3.24 &  0.00, 3.56 &  0.00, 4.18 \\
   1.7 &  0.72, 2.70 &  0.38, 3.34 &  0.06, 3.66 &  0.00, 4.28 \\
   1.8 &  0.81, 2.80 &  0.45, 3.44 &  0.16, 3.76 &  0.00, 4.38 \\
   1.9 &  0.90, 2.90 &  0.51, 3.54 &  0.26, 3.86 &  0.00, 4.48 \\
   2.0 &  1.00, 3.00 &  0.58, 3.64 &  0.35, 3.96 &  0.00, 4.58 \\
   2.1 &  1.10, 3.10 &  0.65, 3.74 &  0.45, 4.06 &  0.00, 4.68 \\
   2.2 &  1.20, 3.20 &  0.72, 3.84 &  0.53, 4.16 &  0.00, 4.78 \\
   2.3 &  1.30, 3.30 &  0.79, 3.94 &  0.61, 4.26 &  0.00, 4.88 \\
   2.4 &  1.40, 3.40 &  0.87, 4.04 &  0.69, 4.36 &  0.07, 4.98 \\
   2.5 &  1.50, 3.50 &  0.95, 4.14 &  0.76, 4.46 &  0.17, 5.08 \\
   2.6 &  1.60, 3.60 &  1.02, 4.24 &  0.84, 4.56 &  0.27, 5.18 \\
   2.7 &  1.70, 3.70 &  1.11, 4.34 &  0.91, 4.66 &  0.37, 5.28 \\
   2.8 &  1.80, 3.80 &  1.19, 4.44 &  0.99, 4.76 &  0.47, 5.38 \\
   2.9 &  1.90, 3.90 &  1.28, 4.54 &  1.06, 4.86 &  0.57, 5.48 \\
   3.0 &  2.00, 4.00 &  1.37, 4.64 &  1.14, 4.96 &  0.67, 5.58 \\
   3.1 &  2.10, 4.10 &  1.46, 4.74 &  1.22, 5.06 &  0.77, 5.68 \\
\end{tabular}
\end{table}
} 

{\mediumtext
\begin{table}
\caption{Properties of the proposed technique for setting confidence 
regions in neutrino oscillation search experiments and three
alternative classical techniques defined in the text.}
\label{tab-Brand-X}
\bigskip
\begin{tabular}{lp{20mm}p{20mm}p{20mm}} \hline
Technique & \raggedright Always gives useful results &
\raggedright Gives proper coverage & Is powerful \\ \hline
Raster Scan & &\multicolumn{1}{c}{$\surd$}   \\ 
Flip-Flop Raster Scan & & &  \\ 
Global Scan & & &  \multicolumn{1}{c}{$\surd$}  \\ 
Proposed Technique &\multicolumn{1}{c}{$\surd$} & 
\multicolumn{1}{c}{$\surd$}&\multicolumn{1}{c}{$\surd$}
\end{tabular}
\end{table}
} 

\narrowtext
\begin{table}
\caption{Experimental sensitivity (defined as the average upper limit that
would be obtained by an ensemble of experiments with the expected
background and no true signal), as a function of the expected
background, for the case of a measurement of a Poisson variable.}
\label{tab-sens}
\bigskip
\begin{tabular}{ddddd} \hline
$\mub$ & \multicolumn{1}{c}{68.27\% C.L.}  
& \multicolumn{1}{c}{90\% C.L.} & \multicolumn{1}{c}{95\% C.L.} 
& \multicolumn{1}{c}{99\% C.L.} \\ \hline
  0.0 & 1.29 & 2.44 & 3.09 & 4.74 \\
  0.5 & 1.52 & 2.86 & 3.59 & 5.28 \\
  1.0 & 1.82 & 3.28 & 4.05 & 5.79 \\
  1.5 & 2.07 & 3.62 & 4.43 & 6.27 \\
  2.0 & 2.29 & 3.94 & 4.76 & 6.69 \\
  2.5 & 2.45 & 4.20 & 5.08 & 7.11 \\
  3.0 & 2.62 & 4.42 & 5.36 & 7.49 \\
  3.5 & 2.78 & 4.63 & 5.62 & 7.87 \\
  4.0 & 2.91 & 4.83 & 5.86 & 8.18 \\
  5.0 & 3.18 & 5.18 & 6.32 & 8.76 \\
  6.0 & 3.43 & 5.53 & 6.75 & 9.35 \\
  7.0 & 3.63 & 5.90 & 7.14 & 9.82 \\
  8.0 & 3.86 & 6.18 & 7.49 & 10.27 \\
  9.0 & 4.03 & 6.49 & 7.81 & 10.69 \\
  10.0 & 4.20 & 6.76 & 8.13 & 11.09 \\
  11.0 & 4.42 & 7.02 & 8.45 & 11.46 \\
  12.0 & 4.56 & 7.28 & 8.72 & 11.83 \\
  13.0 & 4.71 & 7.51 & 9.01 & 12.22 \\
  14.0 & 4.87 & 7.75 & 9.27 & 12.56 \\
  15.0 & 5.03 & 7.99 & 9.54 & 12.90 \\
\end{tabular}
\end{table}

\narrowtext


\begin{references}
\bibitem{unbinned}
Other than convenience in displaying and handling the data, there is
no good reason to bin the data at all.  The bins
can be made so fine that there is a most one measured event in each
bin, so that the binned likelihood reduces to an unbinned likelihood.
\end{references}
\end{document}